\documentclass[a4paper,12pt]{article}
\usepackage{amsthm}
\usepackage{amsmath}
\usepackage{amsfonts}
\usepackage{amssymb}
\usepackage{hyperref}
\usepackage{latexsym}
\usepackage{enumerate}
\usepackage{color}
\usepackage{setspace}
\usepackage{comment}
\usepackage{enumitem}
\definecolor{darkgreen}{rgb}{0,0.6,0}
\definecolor{darkblue}{rgb}{0,0,0.3}
\definecolor{darkred}{rgb}{0.7,0,0}

\newcommand{\de}{\hat{\delta}}

\newtheorem{proposition}{Proposition}[section]
\newtheorem{defi}{Definition}[section]

\newcommand{\be}{\begin{equation}}
\newcommand{\ee}{\end{equation}}
\newcommand{\bea}{\begin{eqnarray}}
\newcommand{\eea}{\end{eqnarray}}
\newcommand{\bse}{\begin{subequations}}
\newcommand{\ese}{\end{subequations}}
\newcommand{\beqa}{\begin{eqnarray}}
\newcommand{\eeqa}{\end{eqnarray}}
\newcommand{\beqar}{\begin{eqnarray*}}
\newcommand{\eeqar}{\end{eqnarray*}}

\newcommand{\ba}{\begin{array}}
\newcommand{\ea}{\end{array}}
\newcommand{\bc}{\begin{center}}
\newcommand{\ec}{\end{center}}

\def\sltr{SL$(2,\mathbb{R})$ }
\def\sltruon{SL$(2,\mathbb{R})\times$U$(1)^N$ }

\makeatletter \@addtoreset{equation}{section}

\makeatletter\renewcommand\section{\@startsection {section}{1}{\z@}%
                                   {-3.5ex \@plus -1ex \@minus -.2ex}
                                   {2.3ex \@plus.2ex}%
                                   {\normalfont\large\bfseries}}
\renewcommand\subsection{\@startsection{subsection}{2}{\z@}%
                                     {-3.25ex\@plus -1ex \@minus -.2ex}%
                                     {1.5ex \@plus .2ex}%
                                     {\normalfont\bfseries}}

\begin{document}


\begin{titlepage}
\hfill
\hfill
\vbox{
    \halign{#\hfil         \cr

         IPM/P-2014/028  \cr
         \today\cr }
      }
\vspace*{4mm}

\begin{center}
{\Large{\bf{Near Horizon Extremal Geometry Perturbations:}}}
\vskip3mm

\centerline{\large{\bf{\emph{Dynamical Field Perturbations vs. Parametric Variations}}
}} \vspace{12mm}

{\large{\bf{K. Hajian$^{\dagger ,\ast}$\footnote{kamalhajian@physics.sharif.edu}, A. Seraj$^\dagger$\footnote{ali\_seraj@ipm.ir}, M. M. Sheikh-Jabbari$^\dagger$\footnote{jabbari@theory.ipm.ac.ir} }}}
\\

\vspace{10mm}
\normalsize
\begin{center}
$^\dagger$\textit{School of Physics, Institute for Research in Fundamental
Sciences (IPM), P.O.Box 19395-5531, Tehran, Iran}\\

$^\ast$\textit{Department of Physics, Sharif University of Technology, \\P. O. Box 11155-9161, Tehran, Iran}
\medskip
\end{center}

\end{center}
\setcounter{footnote}{0}

\begin{abstract}
In \cite{NHEG-Mechanics} we formulated and derived the three universal laws governing Near Horizon Extremal Geometries (NHEG). In this work we focus on the Entropy Perturbation Law (EPL) which, similarly to the first law of black hole thermodynamics, relates  perturbations  of the charges labeling perturbations  around a given NHEG to the corresponding  entropy perturbation. We show that  field perturbations governed by the linearized equations of motion and  symmetry conditions which we carefully specify, satisfy the EPL. We also show that these perturbations are limited to those coming from difference of two NHEG solutions (i.e. variations on the NHEG solution parameter space). Our analysis and discussions  shed light on the ``no-dynamics'' statements of \cite{No-dynamics-1,No-dynamics-2}.
\end{abstract}

\end{titlepage}

\setlength{\baselineskip}{1.1 \baselineskip}
\addtocontents{toc}{\protect\setcounter{tocdepth}{2}}
\tableofcontents
\section{Introduction}

It is well known that black holes obey laws of thermodynamics. A symmetry based covariant approach to derivation of the laws of black holes mechanics was introduced in \cite{Wald:1993nt,Iyer:1994ys}. In this approach, entropy and other extensive thermodynamic  parameters of a  black hole are shown to be the Noether-Wald conserved charges associated with the symmetries of the black hole solution. Specifically, entropy is the conserved charge corresponding to the generator of black hole horizon, which is a Killing vector constructed from the Killing symmetries of the geometry and becomes null at the horizon. This relation between the symmetries, leads to a relation between  perturbations in the conserved charges, the ``first law of black hole thermodynamics".

The Noether conserved-charge based approach has two remarkable features: 1) It gives a universal proof of the first law of black holes in any generally covariant theory of gravity in any dimension. 2) It provides a different interpretation and meaning to the first law of black holes than was initially proposed in \cite{BCH}, where the perturbations/variations appearing in the first law are viewed as perturbations/variations in the parameters space of family of black hole solutions. In the Noether-Wald approach \cite{Iyer:1994ys}, however, the charge variations in the first law are attributed to generic perturbations (probes) on a given black hole background.  Derivation in \cite{Wald:1993nt,Iyer:1994ys} asserts that the perturbations, probe fields, which satisfy linearized equations of motion on the background black hole geometry with appropriate boundary conditions, are in thermal equilibrium with  the thermal bath of  background black hole geometry,  which specifies the non-extensive quantities (like temperature and chemical potentials) appearing in the first law. In particular, one can associate entropy to these probes (as well as to the black hole background \cite{Wald:1993nt}).

A crucial assumption in the Wald's approach to the first law is that it can only be applied to geometries with a Killing horizon, this assumption is generically fulfilled by stationary black holes.  Moreover, it requires the Killing horizon to be a bifurcate horizon, i.e. the black hole should necessarily have a non-vanishing temperature. The existence of bifurcate horizon is required, as entropy (and its perturbations) are defined as integrals over the codimension two bifurcation surface and the corresponding Killing vector is normalized by surface gravity of the black hole. The question which then arises naturally is the existence of a relation between conserved charges and their perturbations/variations, of \emph{extremal} black holes  which have zero surface gravity (Hawking temperature) and no bifurcation horizon.

Even if one assumes that the general form of first law is valid for extremal black holes (e.g. using the physical expectation that the first law should be continuous in its parameters and in particular temperature), the first law at zero temperature reduces to a manifestation of extremality (BPS) relation and does not determine the entropy perturbations in terms of other charge perturbations, simply because  perturbation of the entropy is not present due to the vanishing of temperature. Through a careful analysis of vanishing temperature limit of the first law together with generic properties of near extremal black holes the ``entropy perturbation law" for extremal black holes was obtained which relates variations/perturbations of the entropy for extremal black holes to perturbations/variations of its other charges \cite{First-law}.

A more concrete derivation of the Entropy Perturbation Law (EPL) for extremal black holes was presented in \cite{NHEG-Mechanics}, carrying out steps similar to Wald's derivation \cite{Iyer:1994ys} for the Near Horizon Extremal Geometries (NHEG). In \cite{NHEG-Mechanics}, we focused on the NHEG as family of solutions to gravity theories (independently of extremal black holes) and showed that despite the absence of Killing or event bifurcate horizon, one can still define an entropy as a conserved Noether-Wald charge of this space through integration of the appropriate entropy density two-form over a codimension two surface which can be unambiguously defined using the \sltr isometry of the NHEG background. Using this approach we derived the ``entropy law'' which is a universal  relation between the entropy and other conserved Noether-Wald charges associated with the NHEG. The entropy law is specific to NHEG and has no counterpart in the usual black hole mechanics.

As mentioned above, in \cite{NHEG-Mechanics} we also derived an entropy perturbation law for NHEG. While in the derivation of entropy law we could completely rely on the \sltr invariance of the NHEG background for defining the two-form conserved charge densities and the integration surface,  the perturbations which satisfy EPL are not generically  \sltr invariant and this may introduce a dependence on the integration surface for the charge perturbations appearing in the EPL. In this paper we revisit the derivation of EPL, paying special attention to this feature and show that one can conveniently derive the EPL which is independent of the surface of integration defining the charges, if we restrict the field perturbations to respect a part of \sltr invariance of the background.  We will argue that this restriction is very well justified when we consider the extremal black hole leading to the NHEG in question in its near horizon limit.

We then study which field perturbations satisfy the conditions required in the derivation of EPL (these conditions are linearized field equations and invariance under the two dimensional subgroup of SL$(2,\mathbb{R})$). Adding appropriate/necessary ``boundary conditions" to these two conditions we show that these perturbations are uniquely determined by their charges and can only be the perturbations which relate to nearby NHEG solutions in the parameters space of NHEG solutions. Our analysis here provides a new viewpoint on, as well as an extension of, the results of \cite{No-dynamics-1,No-dynamics-2} where a ``no dynamics" theorem in near horizon extremal Kerr (NHEK) geometry was presented. Our uniqueness theorem opens a new way of studying boundary gravitons and the possibility of identification of microstates giving rise to extremal black hole or the corresponding NHEG entropy.

Organization of this paper is as follows. In section 2, we will give a brief review of NHEG geometry and its universal laws. In section \ref{sec NHEG dynamic}, we summarize the conditions defining  NHEG dynamical field perturbations and conditions for the entropy perturbation law be independent of surface of integration over which the charge perturbations are defined. In section \ref{sec NHEG parametric}, we show that field perturbations which correspond to the difference of two NHEG solutions satisfy the conditions defining dynamical field perturbations discussed in section \ref{sec NHEG dynamic}, and that these perturbations satisfy the EPL.
In section \ref{sec uniqueness}, we present the NHEG perturbations uniqueness theorem: The only field perturbations which satisfy the three conditions defining dynamical field perturbations outlined in section \ref{sec NHEG dynamic}, are those discussed in section \ref{sec NHEG parametric} which correspond to the variations in the family of NHEG solutions.
In the last section we summarize our results and make concluding remarks. In three appendices we have gathered some more technical details of the computations.

\section{Review of NHEG's and three laws of NHEG mechanics}

Near Horizon Extremal Geometries (NHEG) are a generic family of solutions to (Einstein-Maxwell-Dilaton, EMD for short) gravity theory. As their name suggests, they have been first obtained and studied in connection with extremal black holes and their near horizon limit \cite{Sen-entropy-function, NHEG-general,NHEG-FKLR,NHEG-KL, NHEG-FL,NHEG-KL-review}. Given the metric of a stationary extremal black hole in $d$ dimensions, with $n$ axisymmetric coordinates and $N\!-\!n$ U$(1)$ gauge fields (producing U$(1)^N$ symmetry) and arbitrary numbers of dilaton fields, one can apply the near horizon limit which is  a specific coordinate transformation  associated with near horizon expansion, accompanied by an appropriate scaling and limit, to obtain the NHEG.

One can present NHEG metric by coordinates in which the SL$(2,\mathbb{R})\times$U$(1)^n$ symmetry is manifest:
\begin{align}\label{NHEG metric}
ds^2=\Gamma\left[-r^2dt^2+\dfrac{dr^2}{r^2}+\sum_{\alpha, \beta=1}^{d-n-2} \Theta_{\alpha \beta}d\theta^\alpha d\theta^\beta+\sum_{i,j=1}^n\gamma_{ij}(d\varphi^i+k^irdt)(d\varphi^j+k^jrdt)\right]\,,
\end{align}
and a set of gauge fields $A^{(p)}$
\begin{align}\label{eq gauge fields}
A^{(p)}=\sum_{i=1}^n f^{(p)}_i(d\varphi^i+k^irdt)+e^prdt\,,
\end{align}
and dilatons:
\begin{equation}
\phi^I=\phi^I(\theta^\alpha)\,,
\end{equation}
 where $i,j=1,\cdots, n$ ($n\leq d-3$),  $p=n+1,\cdots, N$,   and $I$ counts arbitrary number of dilatons. $\Gamma,\Theta_{\alpha \beta},\gamma_{ij},f^{(p)}_i$ are functions of the polar coordinates $\theta^\alpha$ whose explicit form can be fixed using equations of motion. The constant $t,\ r$ surfaces in the metric \eqref{NHEG metric}, which are spanned by $\theta^\alpha, \varphi^i$ are chosen to be smooth and compact (finite volume) $d-2$ dimensional surfaces. The fields in the NHEG solution, namely metric $g_{\mu\nu}$, gauge fields $A^{(p)}$ and dilatons $\phi^I$ will be collectively denoted by $\Phi$.

NHEG's have some generic features  \cite{NHEG-Mechanics,{NHEG-KL-review}}:
\begin{itemize}
\item They are solutions to the equations of motion of the same theory as the original extremal black holes were and hence establish a new  independent family of solutions. Unlike the original extremal black hole, the NHEG is not asymptotic to a maximally symmetric geometry and also has not an event horizon.
\item NHEG's have time-like Killing vector field, and may hence be regarded as a stationary geometry. This time-like Killing vector field, however, is not generically or necessarily globally defined \cite{Compere-review}.
\item They have  an AdS$_2$ factor and accordingly an SL$(2,\mathbb{R})$ symmetry.
\item It inherits the U$(1)^N$ symmetry from the extremal black hole; the NHEG solution has then SL$(2,\mathbb{R})\times$U$(1)^N$  symmetry.
\item In the above coordinates,  Killing vectors generating the SL$(2,\mathbb{R})\times$U$(1)^n$ symmetry are:
\begin{align}\label{sl2r-generators-xi-a}
\xi_1 &=\partial_t\,,\cr
\xi_2 &=t\partial_t-r\partial_r\,,\\
\xi_3 &=\dfrac{1}{2}(t^2+\dfrac{1}{r^2})\partial_t-tr\partial_r -\sum_{i=1}^n\dfrac{k^i}{r}\partial_{\varphi^i}\,,\cr
m_i &=\partial_{\varphi^i}\,,
\end{align}
with the commutation relations:
\begin{align}\label{SL2R}
\left[\xi_1,\xi_2\right]&=\xi_1\,, \quad \quad \left[\xi_2,\xi_3\right]=\xi_3\,,  \quad \quad \left[\xi_1,\xi_3\right]=\xi_2\,,\\
\left[\xi_a,m_i\right]&=0\,,\quad  a\in\{ 1,2,3\} \;\ \text{and},\ \  i\in \{1,\dots , n\}\,.
\end{align}
\item For $n=d-3$ there are uniqueness theorems (see \cite{NHEG-KL-review} for a review). Therefore, the geometry is \emph{uniquely} determined  by $N$ conserved charges associated with U$(1)^N$ symmetry. That is, $n$ angular momenta $J_i,\quad i\in \{1,2,\dots,n\}$ and $N-n$ electric charges $q_p,\quad p\in \{n+1,\dots,N\}$. There could also be $N-n$ magnetic charges, which are generically topological, and not Noether charges and hence do not directly appear in  our analysis and their presence will not change our results.

\item In NHEG's, there are two independent vector fields which are null on the whole geometry. In Poincar\'{e} coordinates \eqref{NHEG metric}, they are
\be\label{ell-wp-null}
\begin{split}
\ell^\mu&=(\dfrac{1}{r^2},1,0,-\dfrac{k^i}{r}),\\
n^\mu&=\dfrac{r^2}{2\Gamma}(\dfrac{1}{r^2},-1,0,-\dfrac{k^i}{r}),
\end{split}
\ee
the normalization is chosen such that $\ell\cdot n=-1$. Note that
\begin{align}
\ell \cdot \nabla \ell^\mu=0,\hspace{1cm}
n\cdot\nabla n^\mu= \frac{-r}{\Gamma}\, n^\mu\,.
\end{align}
This shows that $\ell,n $ are the generators of two null geodesic congruences ($\ell$, unlike $n$, is affinely parameterized). Therefore, the near horizon geometry is a Petrov type D spacetime. Moreover, these are null geodesics with vanishing expansion, rotation and shear, and hence NHEG is a Kundt spacetime \cite{Durkee:2010ea}.

\item $\ell, n$ vector fields are normal to the vectors $\partial_{\theta^\alpha}$ and $\partial_{\varphi^i}$. Therefore, the binormal to constant $t, r$ surfaces $H$ is
\begin{align}
{\epsilon}_{\mu\nu}&=\ell_{[\mu}n_{\nu]}\,.
\end{align}
The normalization  $\ell\cdot n=-1$ implies that $\epsilon_{\mu\nu}\epsilon^{\mu\nu}=-2$.
\end{itemize}

The idea proposed and analyzed in \cite{NHEG-Mechanics} is studying (thermo)dynamic properties of the NHEG, considering it as an independent solution to a covariant gravity theory (which is chosen to be Einstein-Maxwell-Dilaton, EMD) and is determined by requesting \sltruon symmetry. The motivation for this proposal is twofold: 1) It is widely believed that microstates of an extremal black hole reside somewhere near its horizon, so studying NHEG might open a new insight to the unresolved problem of microstates of black holes. 2) It extends the thermodynamic behavior observed in black hole solutions of gravity theories  to another family of solutions which do not have event horizons.

Although NHEG's do not have any bifurcate Killing horizon or event horizon, constant $t$ and $r$ surfaces, defined at  arbitrary $t=t_H$ and $r=r_H$ in \eqref{NHEG metric}, provide  an infinite set of $d-2$ dimensional  smooth and compact surfaces which may be viewed as the ``Killing horizon'': All these surfaces have the same volume-form which is \sltr invariant \cite{NHEG-Mechanics}. The NHEG's have  a Killing vector $\zeta_H$
\be\label{zetaH}
\zeta_H={n}^a_H\xi_a-k^im_i\,,
\ee
where $n^a_H$ is the unit vector of the SL$(2,\mathbb{R})$,\footnote{The $n^a$ and $n_H^a$ should not be confused with the null vector $n^\mu$ defined in \eqref{ell-wp-null}.}
\begin{align}\label{n-a-r-t}
n^1=-\frac{t^2r^2-1}{2r}\,,\qquad n^2=tr\,,\qquad n^3=-r\,,
\end{align}
computed at $t=t_H,\ r=r_H$. One can readily  check that $\zeta_H$ vanishes at $H$. The NHEG entropy $S$ can hence be defined as the conserved charge associated with $\zeta_H$, as is done in Wald's formulation for black holes \cite{Wald:1993nt}.

The three laws of NHEG mechanics (paralleling those of black hole mechanics \cite{BCH}) are \cite{NHEG-Mechanics}
\begin{enumerate}
\item \textbf{Zeroth law:} The coefficients $k^i$ and $e^p$ are constant, \textit{i.e.} independent of the coordinates $\theta^\alpha$.\footnote{We note that NHEG solution is not necessarily completely or uniquely specified in terms of the $k^i$ and $e^p$. We will discuss this point further in section \ref{sec-delta-two-hat}.}

\item \textbf{Entropy law:} For any given NHEG there is always the following relation:
\begin{equation}\label{entropy law}
{\frac{S}{2\pi}=k^iJ_i+e^pq_p-\oint_H \sqrt{-g}\mathcal{L}}
\end{equation}
in which $\mathcal{L}$ is the Lagrangian density of the theory and $\oint_H \sqrt{-g}\mathcal{L}$ is calculated on an $H$ surface defined at arbitrary $r_H$ and $t_H$.
\item \textbf{Entropy perturbation law:} For the perturbations (probes) around a given NHEG (satisfying some ``appropriate conditions") we have:
\begin{equation}\label{entrop-pert-law}
{{\frac{\delta S}{2\pi}=k^i \delta J_i +e^p\delta q_p\,}}\,.
\end{equation}
\end{enumerate}
The main goal of the next section is introducing and justifying the ``appropriate conditions" for the perturbations of dynamical fields around NHEG leading to the EPL. These conditions will be used to specify these perturbations.

\section{NHEG dynamical field perturbations}\label{sec NHEG dynamic}
In this section we will study  perturbations over NHEG geometries and derive a relation between the charges associated to these perturbations. If we denote the background field configuration of an NHEG solution by $\Phi_0$, we consider field perturbations around this background $\delta\Phi$. This section provides a more precise and detailed definition of dynamical field perturbations $\delta\Phi$ and derivation of the entropy perturbation law given in \cite{{NHEG-Mechanics}}.
\begin{defi}\label{constraints}
Dynamical field perturbations $\delta\Phi$ are defined with  the following properties. That is, $\delta\Phi$
\begin{enumerate}[label=(\Roman*)]
\item \label{condition0} satisfy  linearized field equations,
\item \label{condition1} are stationary and symmetric under scaling, i.e $\mathcal{L}_{\xi_a}\delta\Phi=0,\ \ a=1,2$,
\item \label{condition2} and {asymptotically} respect the isometries of the background. Explicitly \\
$
\lim_{r\rightarrow\infty}\mathcal{L}_{\xi_a}\delta\Phi=0,\ \ a=1,2,3$ \ and \  $\lim_{r\rightarrow\infty}\mathcal{L}_{m_i}\delta\Phi=0 ,\ \ i=1,\cdots ,n$.

\end{enumerate}
\end{defi}
\begin{proposition}\label{EPL-dynamcial-pert}
The charge perturbations corresponding to any field perturbations satisfying  conditions \ref{condition0} and \ref{condition1} satisfy the EPL relation:
\begin{equation}\label{EPL}
{\frac{\delta S}{2\pi}=k^i \delta J_i +e^p\delta q_p}\,.
\end{equation}
\end{proposition}
Proof of the above proposition will be given in section \ref{sec EPL proof}. However, before giving the proof, we discuss physical meaning and justification of the conditions enumerated above.

\subsection{Physical relevance of conditions on $\delta\Phi$}\label{sec EPL conditions}

The fact that field perturbations $\delta\Phi$ should satisfy linearized equations of motion is needed for the (on-shell) conservation of the corresponding Noether-Wald charge densities \cite{NHEG-Mechanics,Iyer:1994ys}. Below, we will discuss requirement of symmetry of perturbations under transformations generated by $\xi_1,\ \xi_2$, i.e. $\mathcal{L}_{\xi_1}\delta\Phi=\mathcal{L}_{\xi_2}\delta\Phi=0$ and the asymptotic symmetry of the perturbations.

\subsubsection{$\xi_1,\ \xi_2$ invariance  of perturbations}\label{entropy pert ind}
 $\xi_1$ is the generator of translations along the time direction of  NHEG geometry $t$ and $\xi_2$ is the generator of scaling
\be\label{scale-transf}
t\to t/k\,,\qquad r\to k r\,,
\ee
in the NHEG metric \eqref{NHEG metric}. Moreover, recalling their Lie-bracket $[\xi_1,\xi_2]=\xi_1$, they form a maximal subgroup of the SL$(2,\mathbb{R})$ isometry group. Below, we provide two arguments for requiring invariance of perturbations $\delta\Phi$ under this subgroup.
One is based on the near horizon limit procedure which relates the NHEG perturbations to perturbations of the associated extremal black hole. The other one follows from the physical requirement that the EPL and all charge perturbations should be independent of the choice of the surface $H$, and that any given point on the AdS$_2$ part of the NHEG metric \eqref{NHEG metric} can be mapped to a point with given $t=t_H,\ r=r_H$ by diffeomorphisms generated by $\xi_1,\ \xi_2$.

\paragraph{Argument 1:} \emph{Perturbations of an extremal black hole  which survive the near horizon limit and  are well-behaved under the limit, give rise to perturbations on NHEG which are invariant under $\xi_1$ and $\xi_2$ diffeomorphisms.}

To see the above consider an extremal black hole with the following metric
\begin{align}
ds^2&=-\tilde{f} d\tau^2+\tilde{g}_{\rho\rho}d\rho^2+\tilde {g}_{\alpha\beta}d\theta^\alpha d\theta^\beta +\tilde{g}_{ij}(d\psi^i-\omega^i d\tau)(d\psi^j-\omega^j d\tau)\,.
\end{align}
It is well known that this geometry has a well defined near horizon limit, defined through the coordinate transformations (e.g. see \cite{NHEG-Mechanics} for more on the conventions and notations)
\begin{align}
\rho &=r_e(1+\lambda r)\,,\qquad
\tau =\dfrac{\alpha r_e t}{\lambda},\hspace{1cm}
\varphi^i =\psi^i-\Omega^i \tau\,,\quad \lambda\to 0\,,
\end{align}
where $r_e$ is the horizon radius, $\Omega^i=\omega^i(r=r_e)$, and $\alpha$ is an irrelevant constant which we can ignore in the computations. Also we set $r_e=1$.

Next, we perturb the extremal black hole geometry $\bar{g}_{\mu\nu}$ by a metric perturbation $\tilde{h}_{\mu\nu}$, that is  the metric for perturbed geometry is $g_{\mu\nu}=\bar{g}_{\mu\nu}+\tilde{h}_{\mu\nu}$. We are searching for perturbations which have a well defined near horizon limit. That is, we are looking for $\tilde{h}_{\mu\nu}$ with  finite $\tilde{h}_{\mu\nu}dx^\mu dx^\nu$ in the near horizon limit.
For the ease of notation let us focus on the 4d case:
\begin{align}
\nonumber \tilde{h}_{\mu\nu}dx^\mu dx^\nu&=\tilde{h}_{\tau\tau}d\tau^2+2d\tau (\tilde{h}_{\tau\theta}d\theta+\tilde{h}_{\tau\psi}d\psi+\tilde{h}_{\tau \rho}d\rho)\\
\nonumber & +\tilde{h}_{\rho\rho}d\rho^2+2d\rho(\tilde{h}_{\rho\theta}d\theta+\tilde{h}_{\rho\psi}d\psi)\\
 &+\tilde{h}_{\theta\theta}d\theta^2+2\tilde{h}_{\theta\psi}d\theta d\psi+\tilde{h}_{\psi\psi}d\psi^2.
\end{align}
Using $d\psi=d\varphi +\Omega d\tau$ and collecting powers of $d\tau=\dfrac{dt}{\lambda}$ and $d\rho=\lambda dr$ yields
\begin{align}
\nonumber \tilde{h}_{\mu\nu}dx^\mu dx^\nu&=\dfrac{d {t}^2}{\lambda^2}\Big(\tilde{h}_{\tau\tau}+2\Omega \tilde{h}_{\tau\psi}+\Omega^2 \tilde{h}_{\psi\psi}\Big)\\
\nonumber & +2\dfrac{d {t}}{\lambda} \Big( \lambda d {r}(\tilde{h}_{\tau\rho}+\Omega \tilde{h}_{\rho\psi})+d {\theta}(\tilde{h}_{\tau\theta}+\Omega \tilde{h}_{\psi\theta})+d {\varphi}(\tilde{h}_{\tau\psi}+\Omega \tilde{h}_{\psi\psi})\Big) \\
\nonumber & +\lambda^2d {r}^2\tilde{h}_{\rho\rho}+2\lambda\, d{r}\Big(\tilde{h}_{\rho\theta}d{\theta}+\tilde{h}_{\rho\psi}d {\varphi}\Big)+\Big(\tilde{h}_{\theta\theta}d {\theta}^2+2\tilde{h}_{\theta\psi}d {\theta} d {\varphi}+\tilde{h}_{\psi\psi}d{\varphi}^2\Big).
\end{align}
Therefore perturbation induced on the NHEG (which we denote by $h_{\mu\nu}$) is
\begin{align}\label{h-components}
\nonumber h_{tt}&=\dfrac{\tilde{h}_{\tau\tau}+\Omega \tilde{h}_{\tau\psi}+\Omega^2 \tilde{h}_{\psi\psi}}{\lambda^2},\hspace{1cm}
\nonumber h_{rr}=\lambda^2 \tilde{h}_{\rho\rho}\\
 h_{tr}&=\tilde{h}_{\tau\rho}+\Omega \tilde{h}_{\rho\psi},\hspace{1cm}
 h_{t\theta}=\dfrac{\tilde{h}_{\tau\theta}+\Omega \tilde{h}_{\theta\psi}}{\lambda},\hspace{1cm}
 h_{t\phi}=\dfrac{\tilde{h}_{\tau\psi}+\Omega \tilde{h}_{\psi\psi}}{\lambda}\\
\nonumber h_{\theta\theta}&=\tilde{h}_{\theta\theta},\hspace{1cm}
\nonumber h_{\varphi\varphi}=\tilde{h}_{\psi\psi},\hspace{1cm}
\nonumber h_{r\theta}=\lambda \tilde{h}_{\rho\theta},\hspace{1cm}
\nonumber h_{r\varphi}=\lambda \tilde{h}_{\rho\psi},\hspace{1cm}
h_{\theta\varphi}=\tilde{h}_{\theta\psi}\,.
\end{align}

Note that these perturbations are solutions to the linearized equations of motion and it is generically expected these perturbations to have an oscillatory time dependence with finite frequencies $\nu$:
\begin{align}\label{BH-modes}
\tilde{h}_{\mu\nu}\sim f(\theta)e^{-i(\nu \tau-m\psi)} (\rho-r_h)^x=f(\theta) e^{i(\frac{\nu-\Omega m}{\lambda}) t}\ e^{im\varphi} (\lambda r)^x\,.
\end{align}
It is argued in \cite{First-law,Kerr/CFT} that $\nu-\Omega m\sim \lambda^2$ and the $\lambda$ dependence comes from the radial dependence of the modes.\footnote{See, however, \cite{Ext-pert-Instability}.} Therefore, we see that $\mathcal{L}_{\xi_1}{h}_{\mu\nu}\sim\lambda\to 0$.

Using \eqref{BH-modes} in \eqref{h-components} and requiring to have finite ${h}_{\mu\nu}$ in the $\lambda \rightarrow 0$ limit, fixes  the $r$ dependence of the perturbations  as:
\begin{equation}\label{hmunu-pert-r-power}
{h}_{\mu\nu}=
\begin{pmatrix}
r^2 & 1 & r & r \\
 & 1/r^2 & 1/r & 1/r\\
  & & 1 & 1 \\
  & & & 1
\end{pmatrix},
\end{equation}
in the $(t,r,\theta,\varphi)$ basis. Note that higher orders of $r$ lead to terms with positive powers of $\lambda$ in ${h}_{\mu\nu}$ so that they disappear in the $\lambda\rightarrow 0$ limit. Also, lower orders of $r$ lead to divergence in  ${h}_{\mu\nu}$ which is excluded. Therefore, \eqref{hmunu-pert-r-power} gives the exact $r$-dependence of components (and not just a leading large $r$ behavior). One may readily check that this $r$-dependence is exactly dictated by the condition $\mathcal{L}_{\xi_2}{h}_{\mu\nu}=0$ (see also \cite{NHEG-general,NHEG-KL-review}). Similar argument may be repeated for the gauge and dilaton fields with a similar conclusion.\footnote{It is instructive to note the similarity and the differences between \eqref{hmunu-pert-r-power} and the Kerr/CFT boundary conditions \cite{Kerr/CFT}.}

To summarize, $\nu-\Omega m^2\sim \lambda^2$ leads to $\mathcal{L}_{\xi_1}{h}_{\mu\nu}\sim\lambda\to 0$, i.e. to time-independence of NHEG perturbations $h_{\mu\nu}$; and the $r$-dependence of NHEG perturbations is fixed by the $\mathcal{L}_{\xi_2}{h}_{\mu\nu}=0$.

\paragraph{Argument 2:} \emph{As discussed, there is an arbitrariness in the choice of the point $t_H, r_H$ defining the surface $H$. It was shown in \cite{NHEG-Mechanics} that the entropy of NHEG, $S$, and its  other charges and hence the entropy law, are independent of the
choice of $H$.  It is hence expected the value of \textbf{charge perturbations}, too, to be independent of $H$. As we will show below, the necessary and sufficient condition for this requirement is  $\mathcal{L}_{\xi_1}\delta\Phi=\mathcal{L}_{\xi_2}\delta\Phi=0$.}

We start our argument by recalling that \cite{NHEG-Mechanics}
\begin{equation}
\frac{S}{2 \pi} = -\oint_H \boldsymbol{{\epsilon}_H} E^{\mu\nu\alpha \beta}\epsilon_{\mu \nu}\epsilon_{\alpha \beta}\,,
\end{equation}
in which
\be\label{E-rank 4}
E^{\mu\nu\alpha \beta}=\frac{\delta \mathcal{L}}{\delta R_{\mu \nu \alpha \beta}}\,
\ee
is a tensor built from the background fields, $\epsilon_{\mu\nu}$ denotes  components of the \sltr-invariant two-form  $\Gamma dt\wedge dr$,
and $\boldsymbol{{\epsilon}_H}$ is the $d\!-\!2$ volume-form of the surface $H$,
\be \label{delta S H}
{\boldsymbol{\epsilon_H}}=\text{Vol(H)}\,{\epsilon}_{\alpha_1,\dots,\alpha_{d-2}}(\,\mathrm{d} x^{\alpha^1}\wedge \dots \wedge \,\mathrm{d} x^{\alpha^{d-2}}),
\ee
where $\epsilon_{\alpha_1,\dots,\alpha_{d-2}}$ is the Levi-Civita symbol defined on surface $H$.

Consider the entropy perturbation associated with dynamical field perturbations $\delta\Phi$ around the NHEG background denoted by field configuration $\Phi_0$:
 \begin{align}
\frac{\delta S}{2 \pi}\bigg|_H = - \oint_H \ \frac{\delta(\boldsymbol{{\epsilon}_H}\,\, E^{\mu\nu\alpha \beta}\epsilon_{\mu \nu}\epsilon_{\alpha \beta})}{\delta\Phi}\bigg|_{\Phi_0}\ \delta\Phi\,.
\label{EPL aux 1}
\end{align}
Next, recall that any two arbitrary $H$ surfaces (defined at different values of $t_H, r_H$) are related by a diffeomorphism generated by $\xi_1,\ \xi_2$. $H$-independence of $\delta{S}$ then means that the integrand should be invariant under such diffeomorphisms. That is,
\be
\mathcal{L}_{\xi_a}\left(\frac{\delta(\boldsymbol{{\epsilon}_H}\,\, E^{\mu\nu\alpha \beta}\epsilon_{\mu \nu}\epsilon_{\alpha \beta})}{\delta\Phi}\bigg|_{\Phi_0}\ \delta\Phi\right)=\frac{\delta(\boldsymbol{{\epsilon}_H}\,\, E^{\mu\nu\alpha \beta}\epsilon_{\mu \nu}\epsilon_{\alpha \beta})}{\delta\Phi}\bigg|_{\Phi_0}\ \mathcal{L}_{\xi_a}(\delta\Phi)=0\,,\quad a=1,2\,,
\ee
where in the second equality we used the fact that background fields $\Phi_0$ are \sltr invariant. The above clearly states that $\mathcal{L}_{\xi_1}(\delta\Phi)=\mathcal{L}_{\xi_2}(\delta\Phi)=0$.

The above reasoning can be readily used for any generic conserved charge of NHEG. Explicitly, consider $\mathcal{Q}\big|_H=\oint_H \boldsymbol{\epsilon}_H \,\mathrm{Q}$, then $\delta{\mathcal{Q}}_H$  caused by the dynamical field perturbations around the NHEG background $\Phi_0$, will be $H$-independent only if the integrand $\boldsymbol{\epsilon}_H \,\mathrm{Q}$ is invariant under $\xi_1,\xi_2$ diffeomorphisms,
\be\label{charge-xi1,2}
\mathcal{L}_{\xi_a}(\boldsymbol{\epsilon}_H \,\mathrm{Q})=\frac{\delta(\boldsymbol{\epsilon}_H \,\mathrm{Q})}{\delta\Phi}\bigg|_{\Phi_0}\mathcal{L}_{\xi_a}(\delta\Phi)=0\,\quad \Longleftrightarrow\quad \mathcal{L}_{\xi_a}(\delta\Phi)=0,\quad a=1,2.
\ee

\subsubsection{Asymptotic isometry of perturbations}
After discussing  physical meaning of condition \ref{condition1}, we now discuss and justify  condition \ref{condition2} which plays the role of boundary conditions for  perturbations. To this end, we first note that in order to find  solutions to the e.o.m of a given theory, some boundary conditions are \emph{usually} needed.\footnote{It may happen that  the symmetry requirements we impose on a solution are so restrictive  that they uniquely specify the solution, without the need for a separate boundary conditions. An  example of such cases is the special class of NHEG for which we have uniqueness theorems \cite{NHEG-KL-review}. In these cases the solution is uniquely determined by requesting \sltruon symmetry and smoothness of $H$ surface.} Boundary conditions can usually be expressed in terms of  asymptotic isometries/symmetries. For instance, one can replace the asymptotic flatness in 4d by requesting asymptotic Poincar\'{e} symmetry. Expressing boundary conditions in terms of the symmetries/isometries has the advantage that they could be presented in a covariant, coordinate independent manner.

In the same spirit, to completely specify solutions  to the linearized equations of motion (l.e.o.m) we need to impose boundary conditions on field perturbations. The most {natural} choice for this boundary conditions is to require the perturbations to respect the symmetries of the NHEG background. This is basically what we have required in \ref{condition2}.\footnote{Another covariant boundary condition, besides \ref{condition2}, is the ``subleading  fall-off boundary condition" (used for instance in the work of Brown-Henneaux \cite{Brown-Henneaux}). For the NHEG one can show that it leads to trivial set of perturbations \cite{Compere-review}.}

As another argument for the boundary conditions for perturbations \ref{condition2}, we recall  discussions of
\cite{No-dynamics-1, No-dynamics-2}, where it is shown that asymptotic \sltruon invariance is  the linearized-stability conditions for linearized perturbations $\delta\Phi$. We will discuss further this requirement in the end of next subsection \ref{sec EPL proof}.

\subsection{Further comments on entropy perturbation law for $\delta\Phi$}\label{sec EPL proof}

The proof of entropy perturbation law under conditions spelled out in proposition \ref{EPL-dynamcial-pert} was given in \cite{NHEG-Mechanics}. In Appendix \ref{app review EPL} we have reviewed the arguments of \cite{NHEG-Mechanics}. As reviewed in the appendix, direct Noether-Wald analysis leads to
\begin{align}\label{entropy-pert-Ea}
\frac{\delta S}{2\pi}\bigg|_H&=k^i\delta J_i+e^{p}\delta q_{p}\bigg|_H+n_H^a\delta \mathcal{E}_a\,,
\end{align}
where we have explicitly put the subscripts $H$ for charges defined as integrals over surface $H$ at $r=r_H$ and $t=t_H$ and the two charges $J_i$ and $\mathcal{E}_a$ are defined as integrals over the space-like surface at $r=\infty$ (\emph{cf.} appendix \ref{app review EPL}).
Here we discuss further  implications of the conditions  \ref{condition1} and \ref{condition2} and show how  condition \ref{condition1} can remove the apparent $H$-dependence in \eqref{entropy-pert-Ea}, and more importantly condition \ref{condition2} yields $\delta\mathcal{E}_a=0$.

In section \ref{entropy pert ind} we showed that  $\frac{\delta S}{2\pi}$ is independent of surface $H$ and hence we may drop subscript $H$ on $\delta S$ term.
As for the angular momentum perturbation $\delta J_i$, we recall its definition \eqref{delta-S-delta-J-integrals},
$$\delta J_i \equiv -\oint_\infty \delta \mathbf{Q}_{m_i}.$$
Since pullback of ${m_i} \! \cdot \! \mathbf{\Theta}$ vanishes over any constant $t,r$ surface on NHEG, one can show that $\delta J_i\big|_\infty$ has the same value once the integral at $r=\infty$ is replaced by any arbitrary $r=r_H$ surface. $\delta q_p$ is also independent of surface $H$. To see this, let us recall definition of the electric charge,
\begin{align}
q_p=-\frac{1}{2} \oint_H \boldsymbol{{\epsilon}_H}\,\, \epsilon_{\mu\nu}\frac{\partial \mathcal{L}}{\partial {F^{^{(p)}}}_{\!\!\mu\nu}}\,.
\end{align}
Due to the argument above \eqref{charge-xi1,2} we deduce $\delta q_p$ is independent of surface $H$.\\
So we can rewrite \eqref{entropy-pert-Ea} as:
\begin{align}\label{EPL with Ea}
\frac{\delta S}{2\pi}-k^i\delta J_i-e^{p}\delta q_{p}=n_H^a\delta \mathcal{E}_a\,.
\end{align}
Since the LHS is independent of $t_H, r_H$, the RHS should also be $r_H$ and $t_H$ independent. Noting that there is no $r_H$ dependence in the $\delta \mathcal{E}_a$ (because it is calculated at infinity) and recalling \eqref{n-a-r-t}, we learn that different powers of $r_H$ should vanish separately. That is,
\begin{equation}\label{rH deduction}
\delta \mathcal{E}_1=0\,,\qquad  t_H\delta \mathcal{E}_2-\delta \mathcal{E}_3=0\,.
\end{equation}
Upon the above conditions, the proof of EPL is complete. In other words, for EPL to hold we need to require \eqref{rH deduction} and   $\delta\mathcal{E}_2$ and $\delta \mathcal{E}_3$ need not vanish independently.

We now show that $\delta\mathcal{E}_2, \delta \mathcal{E}_3$ vanish separately if we consider condition \ref{condition2}, the asymptotic \sltruon invariance. To this end, we recall the fact that $\mathcal{E}_a$ are defined as integrals at infinity, explicitly
\be
t_H\delta \mathcal{E}_2-\delta \mathcal{E}_3= \oint_\infty \boldsymbol{{\epsilon}_H}(t_H\delta{E}_2-\delta{E}_3)= \oint_\infty \boldsymbol{{\epsilon}_H}\mathcal{L}_{\xi_3}(t_H\delta{E}_2-\delta{E}_3)=0\,,
\ee
where $\delta E_2,\delta E_3$ are scalars composed of $(\Phi_0,\delta \Phi,\xi_2)$ and $(\Phi_0,\delta \Phi,\xi_3)$ respectively and are bilinear in $\xi_a$ and $\delta\Phi$. In the second equality above we have used \texttt{i)} (asymptotic) $U(1)^n$ symmetry of $\Phi_0$ and $\delta \Phi$, which implies $\delta E_2,\delta E_3$ are independent of coordinates $\varphi^i$; \texttt{ii)} the explicit form
of $\xi_3$ and that it is independent of $\theta^\alpha$ and does not have any component in direction of $\partial_{\theta^\alpha}$; \texttt{iii)} and
$\mathcal{L}_{\xi_3}(t_H\delta{E}_2-\delta{E}_3)={\xi_3}^\mu\partial_\mu (t_H\delta{E}_2-\delta{E}_3)$. This latter, upon expansion in powers of $r$ implies that $\mathcal{L}_{\xi_3}$ does not change $\theta^\alpha$ dependence of the integrand.
Recalling the \sltr algebra, $\mathcal{L}_{\xi_3}\delta E_3=0$ and $\mathcal{L}_{\xi_3}\delta E_2=-\delta E_3$ asymptotically. Therefore, we learn that
\be
\oint_\infty \boldsymbol{{\epsilon}_H}\mathcal{L}_{\xi_3}(t_H\delta E_2-\delta E_3)=
-\oint_\infty \boldsymbol{{\epsilon}_H}\ t_H\delta E_3= -t_H \delta \mathcal{E}_3=0.
\ee
Since $t_H$ is an arbitrary number, we learn that $\delta \mathcal{E}_3=0$ and hence,
\begin{equation}\label{21}
\delta \mathcal{E}_a=0, \quad \forall \,a.
\end{equation}
We have then shown how all the three conditions \ref{condition0}, \ref{condition1} and \ref{condition2} are essential for vanishing of $\delta{\mathcal{E}}_a=0$, while arriving at the EPL \eqref{EPL}, where each and every term in the EPL is $H$-independent, does not require \ref{condition2}.

Before closing this section we also comment that, as is known from the canonical formulation of general relativity, $\delta\mathcal{E}_a$ are generators of asymptotic gauge transformation $x\rightarrow x+\xi_a$ through the Poisson bracket, under the assumption of integrability, conservation and finiteness of charges \cite{Amsel:2009pu}. If one assumes that a symplectic current exists  such that these assumptions are satisfied, then:
\begin{align}
\left[\delta\mathcal{E}_a,\Phi\right]&= \mathcal{L}_{\xi_a}\delta\Phi\big|_{r\rightarrow\infty}\,.
\end{align}
Therefore the condition $\delta\mathcal{E}_a=0$ is equivalent to the statement that $\xi_a , a=1,2,3$ are the \textit{asymptotic symmetries} of dynamical field perturbations
\begin{align}
 \mathcal{L}_{\xi_a}\delta\Phi\big|_{r\rightarrow\infty}=0 ,\hspace{1cm}a=1,2,3\,.
\end{align}

\section{NHEG parametric perturbations}\label{sec NHEG parametric}

In this section we consider a specific set of perturbations around a given NHEG which are produced through moving in the parameter space of NHEG solutions. These perturbations will hence be called \emph{parametric perturbations}. An NHEG
is specified by a set of conserved charges, angular momenta $J_i$ and electric charges $q_p$.\footnote{Note that, as we will discuss in section \ref{sec-delta-two-hat},
specification in terms of charges is more precise than the one in terms of ``conjugate parameters'' $k^i$ and $e^p$. Note also that NHEG uniqueness theorems has been sofar proven for a subset of all NHEG's \cite{NHEG-KL-review} and there might be NHEG's which are not uniquely specified by their conserved charges.}
One may hence denote an NHEG solution by fields $\Phi_{\{J_i,q_p\}}(x)$. A parametric perturbation, denoted by $\de\Phi$, is defined as
\begin{align}
\de \Phi&\equiv\dfrac{\partial\Phi_{\{J_i,q_p\}}}{\partial J_i}\delta J_i+\dfrac{\partial\Phi_{\{J_i,q_p\}}}{\partial q_p}\delta q_p\,.
\end{align}

We start our analysis of parametric perturbations $\de\Phi$ by showing that they indeed fulfill the three conditions stated in the definition \ref{constraints}.
\begin{itemize}\item\textbf{Linearized equations of motion.}  $\de\Phi$ is the difference between two adjacent solutions of field equations, and the conserved charges $J_i$ and $q_p$ do not appear in the equations of motion. Therefore, one can readily deduce that $\de\Phi$ solves the linearized field equations.
\item\textbf{$\xi_1, \xi_2$ invariance.}
The Killing vectors $\xi_1,\xi_2$, and also $m_i$, do not involve any parameters of the NHEG solution (like $k^i$ and $e^p$). Therefore, if $\de\Phi=\Phi_0'-\Phi_0$,
\begin{align}
\mathcal{L}_{\xi}\de\Phi=\mathcal{L}_{\xi}\Phi_0'-\mathcal{L}_{\xi}\Phi_0=0,\hspace{1cm}\xi=\{\xi_1,\xi_2,m_i\}\,.
\end{align}
So, parametric perturbations $\de\Phi$ are not only $\xi_1,\xi_2$ invariant, but also $m_i$ invariant.

We also note that parametric perturbations preserve the null vectors fields $\ell,n$ \eqref{ell-wp-null}, i.e. $\de(\ell^2)=\de(n^2)=0$. Among other things, this also implies that parametric perturbations, too, preserve constant $t,r$ surfaces $H$.

\item\textbf{Asymptotic \sltruon invariance.} The Killing vector $\xi_3$ \eqref{sl2r-generators-xi-a} involves $k^i$ and hence $\de\Phi$ are not in general invariant under \sltruon symmetry of the NHEG. Nonetheless, the $k^i$ dependence of $\xi_3$ is such that $\de\Phi$ are asymptotically $\xi_3$ invariant.
To see this explicitly, let us denote the corresponding Killing vectors of two NHEG solutions $\Phi_0, \Phi'_0$ by $\xi_3$ and $\xi_3'$. Therefore,
\begin{align}\label{Uniqueness 2}
\mathcal{L}_{\xi_3}\Phi_0=\mathcal{L}_{\xi_3'}\Phi'_0=0 \implies \mathcal{L}_{\xi_3}\de\Phi=-\mathcal{L}_{\de\xi_3}\Phi_0,
\end{align}
where $\de\xi_3=\frac{-\de k^i}{r}m_i$, and $\xi_3$ is not the symmetry of $\de\Phi$. However, since $\de\xi_3=\frac{-\de k^i}{r}m_i$, one can see that $\mathcal{L}_{\de\xi_3}\Phi_0\sim \mathcal{O}(1/r^n)$,\ $n\geq 1$, i.e.  $\xi_3$ is an asymptotic symmetry of $\de\Phi$.

To complete the above argument we need to discuss the cases involving gauge fields separately. For the gauge fields $\mathcal{L}_{\xi_3}{A^{ (p)}}$ is not zero, it is a pure gauge transformation:
$$\mathcal{L}_{\xi_3}{A^{ (p)}}=d\left(\frac{e^{ p}}{r}\right)\,, \qquad \mathcal{L}_{\xi_3^\prime}{A^{\prime (p)}}=d\left(\frac{e^{\prime p}}{r}\right)\,,\qquad \mathcal{L}_{\xi_3}\hat{\delta}A^{(p)}=-\mathcal{L}_{\delta \xi_3}A^{(p)}-\frac{\de e^p}{r^2}dr.$$
Hence, gauge fields also exhibit asymptotic $\xi_3$, and hence \sltruon invariance.

\end{itemize}

\subsection{Proof of entropy perturbation law for parametric perturbations}
So far we have introduced two classes of field perturbations, ``dynamical field perturbations'' and ``parametric field perturbations''.
While  dynamical field perturbations act only on dynamical fields (governed by  field equations), parametric perturbations act both on dynamical and nondynamical parameters of an NHEG solution. For example, dynamical field perturbations do not affect the Killing vectors of the background NHEG.
Despite the fact that parametric perturbations fulfill the three conditions of definition \ref{constraints}, our derivation and proof of EPL, reviewed in appendix \ref{app review EPL} and discussed in section \ref{sec NHEG dynamic}, does not immediately extend over the parametric perturbations. This is due to the fact that in the derivation of EPL, we have assumed  the perturbations do not affect the Killing vectors associated to the background geometry.
This was explicitly used in the derivation of EPL for dynamical field perturbations, \emph{cf.} \eqref{eq xi L0}. We should hence
revisit  derivation of the EPL for parametric perturbations. This is the task of this subsection.

Consider a dynamical perturbation $\delta\Phi$ and a parametric perturbation $\de\Phi$ with the same dynamical content, i.e. $\de\Phi=\delta\Phi$.
As noted above,  charge perturbations corresponding to these perturbations can in principle be different. However, we will show below that this is not the case. To investigate this, we note that parametric perturbation of the charge associated to a Killing $\xi$ can be expressed as
\begin{align}\label{delta hat vs delta}
\de \mathcal{Q}_\xi = \oint \mathbf{Q}_{\xi'}(\Phi'_0) -\oint \mathbf{Q}_\xi(\Phi_0)
=\delta \mathcal{Q}_\xi +\mathcal{Q}_{\de\xi}\,,
\end{align}
where
\begin{align}\label{nondynamical delta}
\delta \mathcal{Q}_\xi \equiv \oint \mathbf{Q}_\xi(\Phi'_0)-\oint \mathbf{Q}_\xi(\Phi_0)
\end{align}
is the charge perturbation associated with ``dynamical field perturbations'' used in section \ref{sec NHEG dynamic}, and
in its definition, unlike $\de\mathcal{Q}_\xi$, we do not vary the Killing vector.
Since $\de m_i=0$, \eqref{delta hat vs delta} implies that
\begin{align}
\de J_i =\delta J_i\,.
\end{align}
Recalling the definition of electric charges $q_p$, and that it does not involve  non-dynamical fields (such as a Killing vector) we readily have
\be
\de q_p=\delta q_p\,.
\ee
Next, we consider parametric variations of the entropy $\de S$, which using \eqref{delta hat vs delta}  can be written as
\begin{align}\label{deltaS vs delta hat S}
\de S= \delta S+\oint_H \mathbf{Q}_{\de\zeta_H}\,,
\end{align}
where
\begin{align}
\de \zeta_H=-\de k^i(\dfrac{r_H}{r}-1) m_i\,.
\end{align}
According to Wald's decomposition theorem \cite{Iyer:1994ys}, one can write the Noether charge corresponding to any diffeomorphism $\zeta$ in the form
\begin{align}
\mathcal{Q}_\zeta &=\oint d\Sigma_{\mu\nu}Q^{\mu\nu}_\zeta
\end{align}
where
\begin{align}
Q^{\mu\nu}_\zeta &= W^{\mu\nu\alpha} \zeta_\alpha -2E^{\mu\nu\alpha\beta}\nabla_\alpha \zeta_\beta +Y^{\mu\nu}+(dZ)^{\mu\nu}\,.
\end{align}
In this equation, the last two terms are ambiguities in the definition of charge which are linear in the generator $\zeta$ and, $E^{\mu\nu\alpha\beta}$ is defined in \eqref{E-rank 4}. For the diffeomorphism $\de \zeta_H$, noting the fact that $\left.\de \zeta_H\right\vert_H=0$,  we have
\be
\begin{split}
{S}_{\de \zeta_H} &=-2\oint_H d\Sigma_{\mu\nu}E^{\mu\nu\alpha\beta}\nabla_\alpha ({\de \zeta_H})_\beta \\
\label{Q eta}&=\oint_H \Big( X^{\alpha\beta}\nabla_\alpha ({\de \zeta_H})_\beta\Big) \;\boldsymbol{\epsilon}_H,
\end{split}
\ee
$\boldsymbol{\epsilon}_H$ is the $d\!-\!2$ volume form of the surface $H$, and have defined,
\begin{align}
X^{\alpha\beta}=-2\epsilon_{\mu\nu}E^{\mu\nu\alpha\beta},
\end{align}
which is an antisymmetric rank two tensor defined on the background fields, and has symmetries of the background. It can be easily checked that any such
tensor has the following form
\begin{align}
X^{\mu\nu}&=\begin{pmatrix}
 0 & F^{tr}(\theta) & 0 & 0 \\
 -F^{tr}(\theta) & 0 & 0 & r F^{r\varphi^i}(\theta) \\
 0 & 0 & 0 & F^{\theta^\alpha\varphi^i}(\theta) \\
 0 & -r F^{r\varphi^i}(\theta)& -F^{\theta^\alpha\varphi^i}(\theta)& 0
\end{pmatrix}
\end{align}
with arbitrary functions $F$ which only depend on $\theta^\alpha$ and have the condition that
\begin{align}\label{SL2R antisymmetric}
X^{ r\varphi^i}=-k^irX^{ rt}\,.
\end{align}
On the other hand, it can be checked that on the surface $H$ we have
\begin{align}\label{hat EPL 1}
\mathcal{H}_{\alpha\beta}\equiv\nabla_{[\alpha} ({\de \zeta_H})_{\beta ]} =\left\lbrace
\begin{array}{ll}
\mathcal{H}_{tr}&=\Gamma\gamma_{ij}k^i\de k^j \\
\mathcal{H}_{r\varphi^i}&= -\Gamma\dfrac{\gamma_{ij}\de k^j}{r}
\end{array}
\right.
\end{align}
and zero otherwise, with the property
\begin{align}\label{SL2R symmetric}
\mathcal{H}_{rt}= r\sum_i k^i\mathcal{H}_{r\varphi^i}\,.
\end{align}
Using \eqref{SL2R symmetric} and \eqref{SL2R antisymmetric} , we have
\begin{align}\label{hat EPL 2}
S_{\de \zeta_H}=\oint_H\Big( X^{rt}\mathcal{H}_{rt} + X^{r\varphi^i}\mathcal{H}_{r\varphi^i}\Big)\boldsymbol{\epsilon}_H
=0\,,
\end{align}
and therefore \eqref{deltaS vs delta hat S} yields $\de S=\delta S$.

Finally, let us consider $\de\mathcal{E}_a$:
\begin{align}
\de \mathcal{E}_a =\delta \mathcal{E}_a + \oint_\infty (\mathbf{Q}_{\rho_a} - \rho_a\cdot\mathbf{\Theta})
\end{align}
where $\rho_a \equiv\de \xi_a$.
\begin{align}
\rho_1=\rho_2=0, \hspace{1cm}\rho_3 =- \dfrac{\de k^i}{r} m_i\,.
\end{align}
It is clear that  $\rho_a\rightarrow 0$ as $r\rightarrow \infty$ and a similar argument like above implies that at $r \rightarrow \infty$
\begin{align}
 \oint_\infty (\mathbf{Q}_{\rho_a} - \rho_a\cdot \mathbf{\Theta})=0\,,
\end{align}
so $\de \mathcal{E}_a=\delta \mathcal{E}_a$.
In brief, we have shown that
\begin{equation}
\hat{\delta}J_i=\delta J_i\,, \qquad \hat{\delta}q_p=\delta q_p\,, \qquad \hat{\delta}S=\delta S\,, \qquad \de \mathcal{E}_a=\delta \mathcal{E}_a=0\,,
\end{equation}
and consequently,
\begin{equation}\label{EPL for hat}
\frac{\hat{\delta} S}{2\pi}=k^i \hat{\delta} J_i +e^p\hat{\delta} q_p\,.
\end{equation}
That is, EPL also holds  for parametric perturbations.

\subsection{Consistency relation for parametric perturbations}
One of the universal laws of  NHEG's is the \emph{entropy law}, which relates entropy to other charges of NHEG. Considering the background NHEG and its adjacent NHEG (call it $\text{NHEG}'$), used to define the parametric perturbation $\hat{\delta}$,
each of these geometries has its own constraint for their parameters, imposed by the entropy law:
\begin{align}
S&=k^iJ_i+e^p q_p-\oint \sqrt{-g}\mathcal{L}\,,\label{entropy law 1}\\
S'&=k'^iJ'_i+e'^p q'_p -\oint \sqrt{-g'}\mathcal{L}'\,.\label{entropy law 2}
\end{align}
Subtracting the above leads to
\begin{align}
\de S=k^i \de J_i+e^p\de q_p +(J_i \de k^i+q_p \de e^p -\de \oint \sqrt{-g} \mathcal{L} )\,.
\end{align}
Using \eqref{EPL for hat},
\begin{align}\label{subtract}
J_i\de k^i +q_p \de e^p =\de \oint \sqrt{-g} \mathcal{L}\,.
\end{align}
This relation is a \emph{consistency} relation for the NHEG perturbations. One can indeed show that \eqref{subtract}, once viewed as
\be
\frac{\de \oint \sqrt{-g} \mathcal{L}}{\de k^i}=J_i\,,\qquad \frac{\de \oint \sqrt{-g} \mathcal{L}}{\de e^p}=q_p\,,
\ee
is basically (a part of) the equations of motion for the NHEG background ansatz \eqref{NHEG metric}, as is also pointed out in Sen's entropy function formalism \cite{Sen-entropy-function}.

\subsection{Isometry preserving perturbations}\label{sec-delta-two-hat}

As discussed in the opening of this section parametric perturbations, except the $\xi_3$ invariance, keep the rest of \sltruon symmetry of the NHEG background.  Here,  we investigate the  question whether  there are a subset of parametric perturbations (which will be denoted by $\hat{\de}$) preserving the full \sltruon symmetry. To answer this question we start noting that
\be
\de\xi_3=\frac{-\de k^i}{r}m_i\equiv \rho_3\,,\qquad
\mathcal{L}_{\xi_3}\de\Phi=-\mathcal{L}_{\rho_3}\Phi_0\,.
\ee
$\hat{\de}$ perturbations are hence those generated by $\rho_3$'s such that $\mathcal{L}_{\rho_3}\Phi_0=0$. In particular,
\begin{align}
\mathcal{L}_{\rho_3}g_{\mu\nu}=0 \implies\gamma_{ij}\hat{\de} k^j=0, \quad \text{or} \qquad \hat{\de} k^i=0\quad  \forall i.
\end{align}
In the last relation we have used smoothness of metric and $H$ surface and that $\gamma_{ij}$ is a non-degenerate matrix.

The question is then whether $\hat{\de}$ family of perturbations are non-empty. To answer this question, let us first consider NHEG solutions to pure gravity theory. Recalling the basic property of vacuum Einstein equations one can show that
\begin{center}
\textit{$k^i=k^i(J_j)$ is a homogeneous function of order zero.}
\end{center}
This implies that $k^i(J_j)=k^i\big((1+\lambda) J_j\big)$, and therefore
\begin{align}\label{hat hat delta}
\hat{\de} J_i = \lambda J_i
\end{align}
is a direction which leaves $k^i$ invariant. The above dovetails with the fact that if metric $g_{\mu\nu}$ is an NHEG solution to $d$ dimensional pure Einstein gravity with angular momenta $J_i$ and entropy $S$, $\kappa^2 g_{\mu\nu}$ is a different NHEG solution with angular momenta $\kappa^{d-2} J_i$ and entropy $\kappa^{d-2}S$, but with the same set of $k^i$. Note that NHEG are not asymptotically flat or (anti)-de Sitter. The discussion above also implies that
there are $n-1$ independent $k^i$'s for an NHEG with $n$ independent angular momenta.

A similar argument can also be made for the NHEG solutions to $d$ dimensional Einstein-Maxwell-Dilaton theory, where the equations of motion are invariant under $g_{\mu\nu}\to \kappa^2 g_{\mu\nu}$ accompanied by $\ A_\mu\to \kappa A_\mu$. Upon this scalings an NHEG solution with parameters $k^i, e^p$ goes to another NHEG with parameters $k^i, \kappa e^p$ (that is, $\hat{\de} k^i= 0,\ \hat{\de} e^p\neq 0$), while the charges transform as $J_i\to \kappa^{d-2} J_i,\ S\to \kappa^{d-2}S,\ q_p\to \kappa^{d-3} q_p$.

\section{Uniqueness of NHEG perturbations}\label{sec uniqueness}
This section contains our main result which is stated in the following proposition:
\begin{proposition}\label{uniqueness prop}
Perturbations around any given NHEG solution to $d$ dimensional EMD theory with SL$(2,\mathbb{R})\times$U$(1)^{d-3}$ isometry,
subject to the conditions of definition \ref{constraints} and with given  charge perturbations $\delta J_i,\ \delta q_p$, are restricted to
the NHEG parametric perturbations  $\hat{\delta}\Phi$. In other words, the only solution to the EPL subject to the three conditions
of definition \ref{constraints} are parametric perturbations $\de\Phi$.
\end{proposition}
Note that the SL$(2,\mathbb{R})\times$U$(1)^{d-3}$ isometry condition has been imposed, because these are the only NHEG backgrounds for which we have uniqueness theorems \cite{NHEG-KL-review}, and of course
the above mentioned ``NHEG perturbation uniqueness theorem'' holds only when we have a similar uniqueness at the background level.

\paragraph{Idea of the proof.}
In the previous section, we explicitly showed that parametric perturbations $\de\Phi$ satisfy the conditions in the  definition \ref{constraints},
and therefore  $\{\hat{\delta}\Phi\}\subset \{\delta\Phi\}$. So, our proof will be complete if we show that the converse is also true, i.e.
$\{\delta\Phi\} \subset \{\hat{\delta}\Phi\}$. To this end we first parameterize field perturbations and simplify them using the symmetry conditions we have assumed and then impose linearized
equations of motion.

We have given an alternative argument in Appendix \ref{Appendix-C} using the gauge invariant analysis of perturbations proposed first by Teukolsky \cite{Teukolsky:1973ha}.

\subsection{Parameterizing field perturbations}
In the EMD theory we are interested in, there are metric, Maxwell gauge fields and dilatons. Here we discuss their perturbations separately.

\subsubsection{Parametrization of  metric perturbations $\delta g_{\mu\nu}$}
Requiring $\delta g_{\mu\nu}$ to have $\xi_1$ and $\xi_2$ symmetries fixes
$\delta g_{\mu\nu}$ to the form (see analysis of appendix \ref{Uniqueness app 1})
\begin{equation}
\delta g_{\mu\nu}=\begin{pmatrix}r^2h_{tt} &  h_{tr} & rh_{t\theta^\alpha}&rh_{t \varphi^i}\\ & \frac{h_{rr}}{r^2}&\frac{h_{r\theta^\alpha}}{r}& \frac{h_{r \varphi^i}}{r}\\ & & h_{\theta^\alpha\theta^\beta}&  h_{\theta^\alpha\varphi^i}\\ & & &h_{\varphi^i\varphi^j}
\end{pmatrix}
\end{equation}
in which $h_{\mu\nu}=h_{\mu\nu}(\theta^\alpha,\varphi^i)$. Discussions  in the appendix \ref{Uniqueness app 1} imply that
requesting asymptotic SL$(2,\mathbb{R})\times$U$(1)^n$ symmetry makes $ h_{tr}=h_{r\theta^\alpha}=h_{r \varphi^i}=0$ and, that the asymptotic $U(1)^n$ isometry is extended to the whole bulk,
removing the $\varphi^i$ dependence of the remaining $h$'s, except for  $h_{rr}$.

Hereafter, we restrict to cases with U$(1)^{d-3}$ isometry, i.e. to the cases where there is only one $\theta$-type coordinate. In these cases
$h_{\theta \varphi^i}$ can be removed by the diffeomorphism $\varphi^i\rightarrow \varphi^i+ f^i(\theta)$ and $h_{\theta\theta}$ may be removed by the
remaining diffeomorphism $\theta\rightarrow \theta+ g(\theta)$, and therefore,
\begin{equation}\label{Uniqueness 4}
\delta g_{\mu\nu}=\begin{pmatrix}r^2h_{tt} & 0 & rh_{t\theta}&rh_{t \varphi^i}\\ & \frac{h_{rr}}{r^2}& 0 & 0 \\ & &
0 &  0\\ & & &h_{\varphi^i\varphi^j}
\end{pmatrix},
\end{equation}
where $h_{rr}=h_{rr}(\theta,\varphi^i)$ and $h=h(\theta)$ for all the other components. Therefore, imposing $\xi_1,\xi_2$ and asymptotic \sltruon invariance, we remain with $(d-1)(d-2)/2+2$  metric perturbation functions. (Alternatively, one could have removed
$h_{t\theta}, h_{t\varphi^i}$ using $\theta, \varphi^i$ diffeomorphisms and remain with  metric perturbations block diagonal in $t,r$ and $\theta,\varphi^i$ parts along codimention two surface $H$.)  We also note that the parametric NHEG metric perturbations $\de h_{\mu\nu}$ can be brought to the form \eqref{Uniqueness 4} with $h_{t\theta}=0$.

\subsubsection{Parameterizing gauge field perturbations $\delta A^{(p)}_{\mu}$}
Let us denote the $N\!-\!(d-3)$ gauge fields in Einstein-Maxwell theory  by $A^{(p)}$. Symmetry conditions of definition \ref{constraints} for perturbations $\delta A^{(p)}$ then imply that (\emph{cf.} appendix \ref{Uniqueness app 1})
\begin{equation}\label{Uniqueness 12}
\delta A^{(p)}=(rh_t^{(p)},0,h_{\theta}^{(p)},h_{\varphi^i}^{(p)})
\end{equation}
in which $h^{(p)}$'s are only functions of $\theta$. $h^{(p)}_\theta$ are simply removed by gauge transformations $\delta A^{(p)}\to \delta A^{(p)}+\,\mathrm{d} \Lambda^{(p)} (\theta)$, so $\delta A^{(p)}$ can be chosen to be:
\begin{equation}\label{Uniqueness 13}
\delta A^{(p)}=(rh_t^{(p)},0,0,h_{\varphi^i}^{(p)})\,,
\end{equation}
which parameterize $(d-2)(N\!-\!(d-3))$ unknown functions. We note that the parametric gauge field perturbations  have the generic form as \eqref{Uniqueness 13} and that these functions are subject to single-valuedness of the gauge field over the $d-2$ dimensional compact surface $H$.
 Moreover,
\eqref{Uniqueness 13} implies that   $r\partial_\theta\delta F_{tr}=\delta F_{t\theta}$, where $F_{\mu\nu}$ is the gauge field strength. This latter is compatible with the parametric field strength perturbation which satisfy $r\partial_\theta\de F_{tr}=\de F_{t\theta}$.

\subsubsection{Parametrization of dilaton perturbations $\delta \phi^I$}

Finally let us consider the dilaton field perturbations $\delta \phi^I$.
Requesting \ref{condition1} and \ref{condition2} for variations of these fields $\delta \phi^I$ also fixes them via lemma in appendix \ref{Uniqueness app 1} to be $\delta\phi^I=\delta \phi^I(\theta)$.

\subsubsection{Regularity and smoothness conditions}\label{regularity sec}
\paragraph{Metric perturbations:} In the cases with SL$(2,\mathbb{R})\times$ U$(1)^{d-3}$ symmetry constant $t,r$  $H$-surfaces are $d-2$ dimensional Euclidean, compact and smooth geometries which are topologically like a ``solid'' torus. That is, at any constant $\theta$ coordinate we find a $d-3$ dimensional torus. The metric on the $H$ surface is
\be
ds^2_H=\Gamma(\theta)\left[d\theta^2+\gamma_{ij}(\theta)d\varphi^id\varphi^j\right]\,,\qquad \varphi^i\in[0,2\pi]\,.
\ee
The smoothness condition then implies that $\Gamma$ cannot have any zeros. If we denote the eigenvalue of the matrix $\gamma_{ij}$ by $\gamma_i(\theta)$,
they should be such that (1) when one of these eigenvalues vanish, the others remain finite, e.g. if $\gamma_i(\theta=0)=0$, then $\gamma_j(\theta=0)\neq 0\,, j\neq i$; (2) First derivative of $\gamma_i$ should also vanish at $\theta=0$ but its second derivative should remain finite, explicitly, around roots of $\gamma_i$ (assuming its located at $\theta=0$), $\gamma_i=\theta^2+{\cal O}(\theta^3),\, \gamma_j=\gamma_j(0), j\neq i$. Considering the whole geometry, the smoothness  conditions in the basis where $\gamma_{ij}$ is diagonal, and around the root of $ii$ component of metric (at $\theta=0$) take the form:
\be\label{smoothness-conditions}
 \frac{g_{\varphi^i\varphi^i}}{g_{\theta\theta}}\sim \frac{g_{\varphi^i t}}{g_{\theta\theta}}\sim \theta^2\,, \qquad \frac{g_{\varphi^j\varphi^j}}{g_{\theta\theta}}\sim \frac{g_{\varphi^j t}}{g_{\theta\theta}}=\text{finite} \ \ j\neq i\,.
\ee
It may of course happen that $\gamma_i$ has more than one roots. Then, the above conditions should hold for all roots.

The above smoothness condition was for the metric itself. It is then readily seen smoothness conditions \eqref{smoothness-conditions} should also be extended to the metric perturbations allowed by our symmetry requirements given in \eqref{Uniqueness 4}. To see this point it is enough to recall that metric perturbations should be relating two metrics $g_{\mu\nu}$ and $g_{\mu\nu}+\delta g_{\mu\nu}$, while these two metrics are both smooth. In particular, since we have adopted a gauge in which $h_{\theta\theta}=h_{\theta\varphi^i}=0$ then, smoothness implies
\be\label{smoothness-h}
h_{\varphi^i\varphi^j}\big|_{\theta=0}\sim \theta^2\,,\qquad {h_{t\varphi^i}}\sim \theta^2\,,\qquad {\partial_\theta h_{t\varphi^j}}\big|_{\theta=0}= 0\,,\ j\neq i\,,
\ee
where $\theta=0$ is the locus $h_{\varphi^i\varphi^i}$ vanishes. Note that for deriving the behavior of $h_{t\varphi^k}$'s we have used the fact that a constant piece in these $h$'s can be absorbed into a shift in $\varphi^k$, at constant $r=r_H$ on a given $H$ surface.

\paragraph{Gauge field perturbations:} To analyze implications of smoothness on the gauge field we consider its field strength
\begin{equation}\label{gauge fields strength pert}
\delta F^{(p)}=h^{(p)}_t\,dr\wedge dt+r\partial_\theta h^{(p)}_t d\theta\wedge dt+\partial_\theta h^{(p)}_{\varphi^i} d\theta\wedge d\varphi^i\,.
\end{equation}
Requiring absence of forces perpendicular to any one of axis of rotations for a charged particle, leads to $\partial_\theta h^{(p)}_t\sim 0$ and $\partial_\theta h^{(p)}_{\varphi^i}\sim 0$ near the pole on that axis \cite{Bergh}. Also from \eqref{gauge fields strength pert} one can see that adding a constant to $h^{(p)}_{\varphi^i}$ does not change the field strength, i.e. there are gauge freedoms for adding constants to $h^{(p)}_{\varphi^i}$ functions.

\paragraph{Dilaton perturbations:} We note that $\delta\phi^I$ should be smooth and single-valued over the $d-2$ dimensional compact surface $H$, this explicitly means that that regularity at ``the pole'' $\theta=0$ fixes $\partial_\theta\delta \phi^I=0$.

\subsection{Imposing field equations}
Having imposed conditions \ref{condition1} and \ref{condition2} on perturbations, we are now ready to impose condition \ref{condition0}, the linearized equations motion. We need to consider equations of motion for metric, gauge fields and dilaton perturbations.

\subsubsection{Linearized equations of motion}
\paragraph{}Linearized Einstein equations takes the form
\be\label{lin-Ein-eq}
G^{(lin)}_{\mu\nu}=T^{(lin)}_{\mu\nu}\,,
\ee
where the LHS in the Einstein-Hilbert theory is
\begin{equation}\label{EH linearized}
G^{(lin)}_{\mu\nu}=\nabla_\alpha \nabla_\nu (\delta g)^\alpha_{\; \mu}+\nabla_\alpha \nabla_\mu (\delta g)^\alpha_{\; \nu}-\Box (\delta g)_{\mu\nu}-\nabla_\nu \nabla_\mu (\delta g)-g_{\mu\nu}[\nabla_\alpha \nabla_\beta (\delta g)^{\alpha \beta}-\Box (\delta g)],
\end{equation}
and the RHS in a general EMD theory is
\begin{equation}\label{T-linearized}
T^{(lin)}_{\mu\nu}=\frac{\delta T_{\mu\nu}}{\delta A_\alpha^{(p)}}\bigg|_{bg.} \delta A_\alpha^{(p)}+\frac{\delta T_{\mu\nu}}{\delta g_{\alpha\beta}}\bigg|_{bg.} \delta g_{\alpha\beta}+\frac{\delta T_{\mu\nu}}{\delta \Phi^I}\bigg|_{bg.} \delta\Phi^I\,,
\ee
where variations are computed on the NHEG background.

We may now plug the field perturbations discussed in previous subsection into \eqref{lin-Ein-eq}. As expected (\emph{cf}. discussions of appendix \ref{Uniqueness app 1}) the linearized Einstein equation takes the form
\begin{equation}\label{Uniqueness 9}
\begin{pmatrix}r^2E_{tt} &  E_{tr} & rE_{t\theta}&rE_{t \varphi^i}\\
& \frac{E_{rr}}{r^2}&\frac{E_{r\theta}}{r}& \frac{E_{r \varphi^i}}{r}\\
 & & E_{\theta\theta}&  E_{\theta\varphi^i}\\
  & & &E_{\varphi^i\varphi^j}
\end{pmatrix}=0\,.
\end{equation}
The main feature of these equations is that because both background and field perturbations have $\{\xi_1,\xi_2\}$ symmetry,
there is not any $(t,r)$ dependence in the coefficients in $E_{\mu\nu}$ above, nor are any derivatives w.r.t these coordinates.
It can be checked  (using the generic shape of background fields and their perturbations discussed in previous sections) that $E_{tr}\!=\!E_{t\theta}\!=\!{E_{r \varphi^i}}\!=\!E_{\theta\varphi^i}\!=\!0$ leads to $h_{t\theta}=0$ and $\partial_{\varphi^i}h_{rr}=0$, removing the only $\varphi^i$ dependence in equations.\footnote{To arrive at this conclusion we have crucially used the form of gauge and dilaton field perturbations and their contribution to the perturbed energy momentum tensor \eqref{T-linearized}.} Therefore, the above simply means $E_{\mu\nu}=0$  are
second order \emph{ordinary} differential equations in $\theta$. Moreover, these equations  are homogeneous \emph{linear} differential equations for remaining  $h_{\mu\nu}$ and $h^{(p)}_\mu$'s and $\delta \phi^I$,  which are of course only functions of $\theta$. Note that the above are showing only a part of l.e.o.m associated with Einstein equations, and there are other equations for gauge field
and dilaton perturbations which will come next.

\paragraph{}{Linearized gauge field equations} in an EMD theory take the form
\begin{equation}\label{Uniqueness gauge equations}
\nabla_\mu  \delta F^{\mu(p)}_{\,\; \nu}+\frac{\delta(\nabla_\mu  F^{\mu(p)}_{\,\; \nu})}{\delta g_{\alpha\beta}}\delta g_{\alpha\beta}
-\alpha_I F^{\mu(p)}_{\,\; \nu}\bigg|_{bg.} \nabla_\mu \delta\phi^I -\alpha_I \nabla_\mu \phi^I\bigg|_{bg.} \delta F^{\mu(p)}_{\,\; \nu}=0\,,
\end{equation}
where $\delta F=\,\mathrm{d} \delta A$ is field strength of the gauge field perturbations, and $\alpha_I$ are constants associated with dilaton-gauge field coupling, through terms like $e^{-\alpha_I\phi^I} F^2$ for each of the gauge fields in the action. Since background and field perturbations  have  $\{\xi_1,\xi_2\}\times $U$(1)^{d-3}$ isometry,  \eqref{Uniqueness gauge equations} is structurally of the form
\begin{equation}\label{uniqueness gauge differential}
(rE^{(p)}_t,\frac{E^{(p)}_r}{r},E^{(p)}_\theta ,E^{(p)}_{\varphi^i})=0,
\end{equation}
where there is no $(t,r,\varphi^i)$ dependence in coefficients of operators in $E^{(p)}=0$'s above. Removing redundant $r$'s in \eqref{uniqueness gauge differential} we remain with the following system of equations,
\begin{equation}
(E^{(p)}_t,E^{(p)}_r,E^{(p)}_\theta ,E^{(p)}_{\varphi^i})=0,
\end{equation}
where $E^{(p)}_\mu=0$'s are ordinary \emph{linear} second order homogeneous differential equations with $\theta$-dependent coefficients.

\paragraph{}{Linearized dilaton equations} provide  one second order ordinary differential equations per each dilaton perturbations $\delta\phi^I$:
\begin{equation}\label{dilaton-pert-equations}
\Box \delta \phi ^I+\frac{\delta (\Box)\phi^I}{\delta g_{\mu\nu}}\delta g_{\mu\nu}+\alpha_J\delta \phi^J \Box \phi^I-\alpha^I e^{-\alpha_J\phi^J} \delta({F_{\mu\nu}^2})=0 \,.
\end{equation}
These unknowns and equations are added to the system of differential equations in \eqref{lin-Ein-eq} and \eqref{Uniqueness gauge equations}.

\subsubsection{Analyzing the l.e.o.m}\label{Analyzing the l.e.o.m}
As discussed above,  linearized equations for all perturbations reduce to some (at most) second order \emph{ordinary}  differential equations with respect to coordinate $\theta$. Moreover, these equations are \emph{linear} in the perturbation fields.
In addition, there are smoothness conditions which these solutions should also satisfy. Dealing with a set of ordinary linear differential equations, if the equations are all consistent with each other (note that number of equations in a crude counting is more than unknown functions), then the solutions are unique up to initial conditions.

On the other hand, as we discussed, these equations do have a set of smooth and regular solutions, the parametric  perturbations $\de\Phi$.
In other words, as we already pointed out, all parametric perturbations $\de\Phi$ are of the form of
\eqref{Uniqueness 4} and \eqref{Uniqueness 13} and satisfy the corresponding l.e.o.m. 
So it just remains to show that for any chosen initial conditions for a member of the set $\{\delta\Phi\}$, there is a member of $\{\hat{\delta}\Phi\}$ which matches that initial conditions. Then uniqueness of the solutions finishes the proof of  $\{\delta\Phi\} \subset \{\hat{\delta}\Phi\}$.

To this end, we need to investigate the linearized equations more closely. Below, we bring the analysis in sentences and words.
These sentences are of course based  on explicit computations and cross-checks for four and five dimensional cases. We have not added the equations to avoid cumbersome, not so illuminating differential equations.

Let us start with  linearized Einstein equations \eqref{lin-Ein-eq} or \eqref{Uniqueness 9} focusing on
$E_{tr}\!=\!E_{t\theta}\!=\!{E_{r \varphi^i}}\!=\!E_{\theta\varphi^i}\!=\!0$ components of equations. As  mentioned in the previous subsection, these equations lead to $h_{t\theta}=0$ and also removes the $\varphi^i$ dependence of $h_{rr}$.
We next note that $E_{r\theta}$ and $E_{\theta\theta}$ components of equations only involve first order differential equations in $\theta$;
they are ``constraint equations'' among the initial conditions. So from the Einstein equations, we remain with $d(d-3)/2+2$ equations and $d(d-3)/2+2$ metric perturbation unknowns\footnote{These equations are $E_{tt}\!=\!E_{rr}\!=\!E_{t\varphi_i}\!=\!E_{\varphi^i\varphi^j}\!=\!0$ and the unknowns are similar components of $h_{\mu\nu}$'s.}, plus the unknowns of gauge fields and dilatons $h^{(p)}_\mu$ and $\delta \phi^I$.

Similarly, one may consider the gauge field equations \eqref{Uniqueness gauge equations}. Noting the allowed form of gauge field perturbations \eqref{Uniqueness 13}, one can readily see that $r,\theta$ components of linearized equation \eqref{Uniqueness gauge equations} is satisfied leading to no extra constraints. Therefore, the number of unknown gauge field components and the corresponding equations become equal\footnote{They are equations $E^{(p)}_t\!=\!E^{(p)}_{\varphi^i}\!=\!0$ and unknowns $h^{(p)}_t,\ h^{(p)}_{\varphi^i}$.} to $d\!-\!2$ for each U$(1)$ gauge field.

Finally, let us discuss the dilaton field perturbations $\delta\phi^I$, which are again subject to second order ordinary differential equations \eqref{dilaton-pert-equations}, one equation per each $\delta\phi^I$.

As discussed above,  number of dynamical equations and unknowns match. Therefore, a member of $\{\delta \Phi\}$ is \emph{uniquely} determined if the initial conditions (which are twice the number of the unknowns, as we are dealing with \emph{linear second order ordinary} differential equations) are completely specified too. Some of the initial conditions are pre-determined by smoothness conditions, therefore it remains to show that  the remaining initial conditions are either constrained to other ones or can be reproduced by labels $\delta J_i$ and $\delta q_p$:
\begin{itemize}
\item For the metric perturbations, two of the initial conditions (which can be chosen to be $\partial_\theta h_{tt}$ and $\partial_\theta h_{rr}$), are constrained to other ones by $E_{\theta\theta}=E_{r\theta}=0$. Also, we note that one can still use gauge freedom (diffeomorphisms) generated by $\xi=t\partial_t$ to subtract off a constant piece from $h_{tt}$.\footnote{Note that these gauge transformations do not change the $(t, r, \varphi^i)$ structure of $\hat{\delta}$ (or $\delta$)  which has been crucially used in our arguments.} The $d(d-3)+1$
    initial conditions for the other components of metric perturbations are completely fixed by
the $(d-2)^2$ smoothness conditions \eqref{smoothness-h} and importantly by the values of $d-3$ angular momentum perturbations $\delta J_i$.
\item For the gauge fields perturbations, initial condition for $h_t^{(p)}\big|_{\theta=0}$ is fixed by the charges $\delta q_p$ and other initial conditions\footnote{They are $\partial_\theta h_t^{(p)}$, $h_{\varphi^i}^{(p)}$ and $\partial_\theta h_{\varphi^i}^{(p)}$ around the pole $\theta=0$.} are fixed using discussion in subsection \ref{regularity sec}.
\item Dilaton fields in the EMD theory has a shift symmetry $\phi^I\to \phi^I+a^I$ for any constant $a^I$. This removes half of the required initial conditions. Recalling our earlier discussions, the regularity and smoothness provides the other half of initial conditions and hence the solutions for dilaton perturbations are also uniquely specified.
\end{itemize}
To conclude this section,  perturbations of an NHEG with SL$(2,\mathbb{R})\times $U$(1)^{d-3}$ isometry and  requirements \ref{condition0}, \ref{condition1} and \ref{condition2}, with a given set of charge perturbations $\delta J_i, \delta q_p$ are uniquely specified by the smoothness conditions.
On the other hand, we already know one such solution, the parametric perturbations $\de\Phi$. Therefore, we have proved the proposition stated in the beginning of this section. In the Appendix \ref{Appendix-C} we have given an alternative argument for our uniqueness theorem.

\section{Concluding remarks}

In this work we continued the analysis of our earlier paper \cite{NHEG-Mechanics} where we had formulated laws of NHEG mechanics. We focused on the entropy perturbation law and tackled the question: which perturbations on the NHEG can lead to charge perturbations satisfying Entropy Perturbation Law (EPL)? To this end, we focused on a set of NHEG field perturbations which satisfy linearized equations of motion of  the Einstein-Maxwell-Dilaton (EMD) theory, to which NHEG is a solution. Importantly, we focused only on the perturbations which keep $\partial_t$ and $t\partial_t-r\partial_r$ Killing vector fields of the background as well as \emph{asymptotically} keeping the \sltruon symmetry of the NHEG background. In section \ref{sec NHEG dynamic} we gave various justifying arguments for these symmetry assumptions on the perturbations. As discussed in \ref{sec EPL conditions}, these symmetry assumption are required if we want to relate NHEG perturbations to the perturbations of an extremal black hole, yielding to the NHEG in consideration in the near horizon limit.
Therefore, our analysis uncovers a class of perturbations of an extremal black hole which satisfy first law of black hole thermodynamics. Of course, we can only  specify the near horizon behavior of these perturbations from our analysis. It would be interesting to study how these perturbations can be extended to the whole bulk of the extremal black hole.

Our main result in this work is the NHEG perturbation uniqueness theorem. We showed by explicit computations that the NHEG perturbations subject to the three conditions discussed above (\emph{cf}. definition \ref{constraints}) is limited to the NHEG parametric perturbations denoted by $\de\Phi$ discussed in section \ref{sec NHEG parametric}. $\de\Phi$ corresponds to the difference of two NHEG solutions which have slightly different conserved charges than $J_i, q_p$ of the background. We proved our NHEG perturbation uniqueness theorem for a class of NHEG solutions with SL$(2,\mathbb{R})\times$U$(1)^{d-3}$ symmetry. These are the NHEG solutions for which we have background NHEG solutions uniqueness theorems (see \cite{NHEG-KL-review} for a review on NHEG uniqueness theorems). The fact that our proof covers all NHEG's for which the background is unique within the given set of charges, is quite natural. Based on the arguments we gave in our proof, we expect that our NHEG perturbation uniqueness theorem can be extended to possible future extensions to the background NHEG uniqueness analysis. Moreover, in our proof we replaced the $U(1)$ symmetry requirements of NHEG background uniqueness theorems \cite{NHEG-KL-review}, with ``asymptotic $U(1)$'' symmetries. This may also show a way to extend such theorems for other NHEG with possibly less symmetries.\footnote{We would like to thank Harvey Reall for a comment on this point.}

Our uniqueness theorem also dovetails with, and in a sense extends, completes and generalizes  the ``no dynamics'' statements of the NHEK background \cite{No-dynamics-1,No-dynamics-2}. We have proved that NHEG perturbations are only limited to those which change an NHEG to another NHEG (near-by in the parameter space). In other words, NHEG cannot be dynamically excited with perturbations which remain normalizable and asymptotically small compared to the background NHEG.
In light of the above discussion and our uniqueness results, one may then revisit the statement of Kerr/CFT correspondence \cite{Kerr/CFT,{Compere-review}} and explore what is the kinematical and dynamical content of  the chiral 2d CFT proposed to be dual to the NHEG. This is what we will discuss in our upcoming paper and here we just discuss our perspective on the issue \cite{HSS-NHEG-microstates}: We have shown that any field perturbation subject to the two conditions (among three) of definition \ref{constraints} is necessarily an NHEG parametric perturbation which satisfies the EPL and is definitely not among the states identified in Kerr/CFT. The Kerr/CFT perturbations should hence be solutions subject to other conditions (than these three). In particular, one can show that Kerr/CFT perturbations are solution to conditions \ref{condition0} and \ref{condition1}, but not \ref{condition2}, so we need to replace the asymptotic symmetry requirement with something more appropriate.
Moreover, the Kerr/CFT perturbations should all have vanishing entropy and charge perturbations, and hence satisfy EPL in a trivial way. This latter is expected, because Kerr/CFT perturbations should parameterize ``microstates'' accounting for the entropy of a given NHEG.

\section*{Acknowledgements}

We would like to thank Geoffrey Comp$\grave{\text{e}}$re, Mahdi Godazgar, Harvey Reall and Hossein Yavartanoo for the comments. We would like to thank the workshop ``Recent Developments in Supergravity Theories'', June 2014, Istanbul and the ICTP Network Scheme Net-68 for providing the stimulating discussion venue. M.M.Sh-J would like to thank Kyung-Hee University, Seoul, Korea, under the international visiting  scholar program.

\appendix

\section{Review of proof of EPL}\label{app review EPL}

This appendix is a review of the discussions in \cite{NHEG-Mechanics} leading to the entropy perturbation law. Starting from the Noether current corresponding to  the diffeomorphism generated by $\zeta_H$:
\begin{equation}\label{eq standard Noether current}
\mathbf{J}_{\zeta_H}=\mathbf{\Theta} (\Phi,\mathcal{L}_{\zeta_H} \Phi)-\zeta_H \! \cdot \! \mathcal{\mathbf{L}}\,,
\end{equation}
we consider variations in \eqref{eq standard Noether current} associated with $\Phi_0\to \Phi_0+\delta\Phi$:
\begin{equation} \label{eq xi L0}
\delta \mathbf{J}_{\zeta_H}=\delta \lbrack \mathbf{\Theta} (\Phi,\mathcal{L}_{\zeta_H} \Phi)\rbrack -{\zeta_H} \! \cdot \! \delta \mathbf{L}\,.
\end{equation}
We assume that the variations do not alter the quantities attributed to the background. In particular, this means that $\delta\zeta_H, \delta\xi_a, \delta m_i$ are all vanishing (as they do in the case of black holes). In this sense these variations are considered as perturbations or \textit{probes} over the NHEG. Let us start our analysis from the  last term in \eqref{eq xi L0}:
\begin{equation}
\delta \mathbf{L}= \mathbf{E}_i \delta \Phi ^i+\mathrm{d} \mathbf{\Theta}(\Phi_0 , \delta \Phi)\,.
\end{equation}
$\mathbf{E}_i$ is the equation of motion for the field $\Phi^i$. The first term vanishes due to the on-shell condition  and the second term is simplified recalling the identity  ${\xi} \! \cdot    \mathrm{d} \mathbf{\Theta} =\mathcal{L}_{\xi} \mathbf{\Theta}-\mathrm{d}({\xi} \!\cdot\! \mathbf{\Theta})$ which is valid for any diffeomorphism $\xi$, therefore,
\begin{equation}\label{eq xi L}
{\zeta_H} \! \cdot \! \delta \mathbf{L}=\mathcal{L}_{\zeta_H} \mathbf{\Theta}(\Phi_0,\delta \Phi) - \mathrm{d} ({\zeta_H} \! \cdot \! \mathbf{\Theta} (\Phi_0,\delta \Phi))\,.
\end{equation}
Inserting the above into \eqref{eq xi L0} we obtain
\begin{equation}\label{deltaJ}
\delta \mathbf{J}_{\zeta_H}=\boldsymbol{\omega}(\Phi_0,\delta \Phi,\mathcal{L}_{\zeta_H}\Phi)  + \mathrm{d} ({\zeta_H} \! \cdot \! \mathbf{\Theta} (\Phi_0,\delta \Phi))\,.
\end{equation}
where
\begin{align}
\boldsymbol{\omega}(\Phi_0,\delta_1 \Phi,\delta_2\Phi)  \equiv\delta_1 \mathbf{ \Theta} (\Phi_0,\delta_2 \Phi) -\delta_2 \mathbf{\Theta} (\Phi_0,\delta_1 \Phi)
\end{align}
is the \textit{symplectic current} \cite{Wald:1993nt,{Iyer:1994ys}}. The current $\mathbf{J}_{\zeta_H}$ is conserved \emph{on-shell}, i.e $\mathrm{d}\mathbf{J}_{\zeta_H}=0$, so one can associate a conserved charge $d-2$ form $\mathbf{Q}_{\zeta_H}$, $\mathbf{J}_{\zeta_H}=\mathrm{d} \mathbf{Q}_{\zeta_H}$,  to the symmetry generated by $\zeta_H$. Moreover, when the solution is deformed by a perturbation which is a solution to the linearized equations of motion, one can take the variation of the relation $\mathbf{J}_{\zeta_H}=\mathrm{d} \mathbf{Q}_{\zeta_H}$ and arrive at
\begin{align}\label{delta,d}
\delta \mathbf{J}_{\zeta_H}=\delta \mathrm{d} \mathbf{Q}_{\zeta_H}=\mathrm{d} \delta \mathbf{Q}_{\zeta_H}\,.
\end{align}
 Using \eqref{delta,d} in \eqref{deltaJ} yields
\begin{equation}\label{conservation-zetaH}
\boldsymbol{\omega}(\Phi_0,\delta \Phi,\mathcal{L}_{\zeta_H}\Phi)  =\mathrm{d}\Big(\delta  \mathbf{Q}_{\zeta_H}- {\zeta_H} \! \cdot \! \mathbf{\Theta} (\Phi_0,\delta \Phi)\Big)\,.
\end{equation}

We integrate the above ``conservation equation'' over a timelike hypersurface $\Sigma$ bounded between two radii  $r=r_H,\;r=\infty$. The hypersurface $\Sigma$ can be simply chosen as a constant time surface $t=t_H$. Integrating \eqref{conservation-zetaH} over $\Sigma$ then yields:
\begin{align}\label{omega vs Q}
\nonumber\Omega(\Phi_0,\delta \Phi,\mathcal{L}_{\zeta_H}\Phi) &= \oint_{\partial\Sigma}\Big(\delta  \mathbf{Q}_{\zeta_H}- {\zeta_H} \! \cdot \! \mathbf{\Theta} (\Phi_0,\delta \Phi)\Big)\\
&=\oint_{\infty}\Big(\delta  \mathbf{Q}_{\zeta_H}- {\zeta_H} \! \cdot \! \mathbf{\Theta} (\Phi_0,\delta \Phi)\Big)-\oint_{H}\delta  \mathbf{Q}_{\zeta_H}\,,
\end{align}
in which we used the definition of  \textit{symplectic form} associated with $\Sigma$ as
\begin{align}
\Omega(\Phi_0,\delta_1 \Phi,\delta_2\Phi) \equiv \int_\Sigma\boldsymbol{\omega}(\Phi_0,\delta_1 \Phi,\delta_2\Phi)\,,
\end{align}
and in the first line we have used the Stokes theorem to convert the integral over $\Sigma$ to an integral over its boundary $\partial\Sigma$ and in the second line, we used the fact that
\begin{equation}
\zeta_H=n_H^a \xi_a-k^im_i
\end{equation}
vanishes on $H$. Since the charge perturbation $\delta\mathbf{Q}_{\zeta_H}$ is linear in the vector $\zeta_H$, one can expand the first term on RHS of \eqref{omega vs Q}
\begin{align}
\Omega(\Phi_0,\delta \Phi,\mathcal{L}_{\zeta_H}\Phi) &= n_H^a\oint_{\infty}\Big(\delta  \mathbf{Q}_{\xi_a}- {\xi_a} \! \cdot \! \mathbf{\Theta}\Big)-k^i\oint_{\infty}\Big(\delta  \mathbf{Q}_{m_i}- {m_i} \! \cdot \! \mathbf{\Theta}\Big)-\oint_{H}\delta  \mathbf{Q}_{\zeta_H}\,.
\end{align}
$m_i$ is tangent to the boundary surface and hence the pullback of ${m_i}\cdot \! \mathbf{\Theta}$ over the surface $r=\infty$ vanishes. It was shown in
\cite{NHEG-Mechanics} that $\Omega(\Phi_0,\delta \Phi,\mathcal{L}_{\zeta_H}\Phi)=-e^p \delta q_p$, where $q_p$ is the electric charge of the gauge field $A^{(p)}$
\begin{equation}
q_p=-\frac{1}{2}\oint_H \epsilon_{\mu\nu}\frac{\partial \mathcal{L}}{\partial {F^{^{(p)}}}_{\!\!\!\!\!\!\mu\nu}}\,.
\end{equation}
Therefore we arrive at
\be\label{EPL-0.0}
\begin{split}
-e^p \delta q_p &= n_H^a\delta \mathcal{E}_a-k^i\oint_{\infty}\delta  \mathbf{Q}_{m_i}-\oint_{H}\delta  \mathbf{Q}_{\zeta_H}\, ,\\
\end{split}
\ee
where $\delta \mathcal{E}_a$ is the canonical generator of the \sltr symmetry $x\rightarrow x+\xi_a$
\begin{align}\label{Ea-def}
\delta \mathcal{E}_a\equiv\oint_\infty\ (\delta \mathbf{Q}_{\xi_a} -{\xi_a}\cdot\mathbf{\Theta})\,,
\end{align}

As usual to Noether-Wald charges \cite{Wald:1993nt,Iyer:1994ys}, there are ambiguities with definition of charges. These ambiguities were dealt with in \cite{NHEG-Mechanics} where it was shown that
\be\label{delta-S-delta-J-integrals}
\frac{\delta S}{2\pi}=\oint_{H}\delta \mathbf{Q}_{\zeta_H}\,,\qquad \delta J_i=-\oint_{\infty}\delta  \mathbf{Q}_{m_i}\,,
\ee
where $\delta S$ and $\delta J_i$ respectively denote the entropy and angular momenta perturbations.
Plugging these into \eqref{EPL-0.0} we obtain
\be\label{EPL-0.1}
\frac{\delta S}{2\pi}= k^i\delta J_i+e^p \delta q_p+n_H^a\delta \mathcal{E}_a\,.
\ee
Note that $S, q_p$ (and their perturbations) are defined on the surface $H$.

\section{Extension of axisymmetry to the bulk}\label{Uniqueness app 1}
\textbf{Lemma:} \textit{Considering field perturbations  $\delta\phi^I$, $\delta A_\mu$ and $\delta g_{\mu\nu}$ in the definition \ref{constraints}, then $U(1)^n$ isometry of these perturbations is extended to all $r$ and is not limited to asymptotic $r\to\infty$ region.}
\begin{proof} We will consider three different field perturbations separately:

\textit{Dilaton:} $\mathcal{L}_{\xi_1}\delta\phi=\partial_t \delta\phi=0$ which means $\delta\phi$ is independent of $t$. $\mathcal{L}_{\xi_2}\delta\phi=r\partial_r\delta\phi=0$ which means $\delta\phi$  is independent of $r$. Therefore $\delta\phi=\delta\phi(\theta^\alpha,\varphi^i)$. Requesting condition \ref{condition2}, i.e $\lim\limits_{r\rightarrow\infty}\mathcal{L}_{m_i}\delta\phi\big|_\infty=0$, leads to $\delta\phi=\delta\phi(\theta^\alpha)$ as desired.

\textit{Vector:} For a covariant vector $A_\mu$,  $\mathcal{L}_{\xi_1}\delta A_\mu=\partial_t \delta A_\mu+\delta A_\nu\partial_\mu \xi_1^\nu=\partial_t \delta A_\mu=0$, therefore its components are independent of $t$. Also the symmetry $\xi_2$ fixes the $r$ dependence of the components as:
\begin{equation}
\delta A_\mu=(rh_t,\frac{h_r}{r},h_{\theta^\alpha} ,h_{\varphi^i} )
\end{equation}
in which $h$'s are some functions of $(\theta^\alpha,\varphi^i)$.  Now assuming the asymptotic $U(1)^n$ symmetry leads to
\be\begin{split}
\forall i \quad 0=\lim\limits_{r\rightarrow\infty}\mathcal{L}_{m_i}\delta A_\nu\big|_\infty&=m_i^\mu \partial_\mu \delta A_\nu+\delta A_\mu \partial_\nu m_i^\nu\bigg|_\infty
=m_i^\mu \partial_\mu \delta A_\nu\bigg|_\infty = \partial_{\varphi^i}\delta A_\nu\bigg|_\infty\,.
\end{split}
\ee
The above leads to $\partial_{\varphi^i}h_t=\partial_{\varphi^i}h_\theta=\partial_{\varphi^i} h_{\varphi^j}=0$. Then,
asymptotic $\xi_3$ symmetry leads to $h_r=0$. This, together with the the general form of gauge field $\delta A_\mu=(rh_t,0,h_{\theta^\alpha} ,h_{\varphi^i} )$, leads to the result that $\delta A_\mu$ is axisymmetirc everywhere.

\textit{Metric:} Considering metric perturbation $\delta g_{\mu\nu}$ as a symmetric second rank tensor, $\mathcal{L}_{\xi_1}\delta g_{\mu\nu}=0$ leads to independence of all components from $t$. $\mathcal{L}_{\xi_2}\delta g_{\mu\nu}=0$ fixes the $r$ dependence as:
\begin{align}
\delta g_{\mu\nu}&=\begin{pmatrix}r^2h_{tt} &  h_{tr} & rh_{t\theta^\alpha}&rh_{t \varphi^i}\\
& \frac{h_{rr}}{r^2}&\frac{h_{r\theta^\alpha}}{r}& \frac{h_{r \varphi^i}}{r}\\
 & & h_{\theta^\alpha\theta^\beta}&  h_{\theta^\alpha\varphi^i}\\
  & & &h_{\varphi^i\varphi^j}
\end{pmatrix}\label{Uniqueness 3}
\end{align}
in which all of the $h$'s are functions of $(\theta^\alpha,\varphi^i)$.
Now assuming the asymptotic axisymmetry leads to
\begin{align}\label{Uniqueness 7}
0 &=\mathcal{L}_{m_i}\delta g_{\mu\nu}\bigg|_\infty =(m_i^\alpha \partial_\alpha \delta g_{\mu\nu}+\delta g_{\mu\nu} \partial_\nu m_i^\alpha+\delta g_{\mu\nu} \partial_\mu m_i^\alpha)\bigg|_\infty
=\partial_{\varphi^i}\delta g_{\mu\nu}\bigg|_\infty\,,
\end{align}
which shows that all component of $\delta g_{\mu\nu}$ are axisymmetric ($\varphi^i$ independent), except for $h_{rr}, h_{r\theta},  h_{r\varphi^i}$ components which are accompanied by powers of $1/r$.
Assuming asymptotic $\xi_3$ symmetry in \ref{condition2}, i.e $\lim\limits_{r\rightarrow\infty}\mathcal{L}_{\xi_3}\delta g_{\mu\nu}=0$ leads to $ h_{tr}=h_{r\theta^\alpha}=h_{r \varphi^i}=0$. In summary, all remaining components of $h$'s are $\varphi^i$ independent, except  $h_{rr}$. However, in section \ref{Analyzing the l.e.o.m} we have discussed that this component is also axisymmetic as a result of linearized field equations.
\end{proof}

\section{An alternative argument for the uniqueness theorem}\label{Appendix-C}
In this appendix we give an alternative argument for proving  the NHEG perturbation uniqueness proposition. The main point in this approach is that perturbations of metric and gauge fields are gauge dependent quantities. So while one can solve the linearized field equations in a  fixed gauge (this is what we have done in section \ref{sec uniqueness}), a more systematic approach is to work with gauge invariant quantities which contain the information about field perturbations, similarly to what is usually done in cosmic perturbation theory, using the gauge invariant quantities (e.g. see \cite{Mukhanov}).

In the context of Petrov type D spacetimes, Teukolsky \cite{Teukolsky:1973ha} introduced a set of gauge invariant scalars built from Weyl tensor and  used them to discuss perturbations of Kerr geometry in a series of papers \cite{Press:1973zz,Teukolsky:1974yv}. It was shown that stability of black hole, interaction with gravitational/electromagnetic waves, and superradiance effects could be studied using these scalars. Teukolsky formulation is based on the Newman-Penrose tetrad \cite{Newman-Penrose}, and the corresponding directional derivative and spin coefficients, which are briefly explained below.

The basic vectors of Newman-Penrose tetrad are the four null vectors $\ell,n,m, m^*$ with the following properties
\be\begin{split}
\ell^2=n^2=m^2={m^*}^2=0\,,\\
\ell \cdot n=-1,\quad m.m^* =1\,.
\end{split}\ee
In the NHEK geometry \eqref{NHEG metric} in four dimensions, the $\ell, n, m$ vectors are explicitly:
\begin{align}
\ell&=
\dfrac{1}{r^2}\partial_t+\partial_r-\dfrac{k}{r}\partial_\varphi\,, \label{null ell}\\
n&=
\frac{r^2}{2 \Gamma (\theta )}\left( \dfrac{1}{r^2}\partial_t-\partial_r-\dfrac{k}{r}\partial_\varphi\right) \,,\\
m&=\frac{1}{\sqrt{2\Gamma(\theta )}}\left( \partial_\theta+\frac{i}{\gamma (\theta )}\partial_\varphi\right)\,.
\end{align}
Using these vectors we can define directional derivative operators
\be\begin{split}
D&=\ell^\mu\nabla_\mu,\qquad \Delta=n^\mu\nabla_\mu,\\
\delta&=m^\mu\nabla_\mu,\qquad \bar{\delta}=m^{*\mu}\nabla_\mu
\end{split}\ee
and construct the spin coefficients \cite{No-dynamics-2}
\begin{align}
\kappa&=-\ell_{a;b}m^a\ell^b  &\nu=n_{a;b}m^{*a}n^b\cr
\rho&=-\ell_{a;b}m^am^{*b} &\mu=n_{a;b}m^{*a}m^b\cr
\sigma&=-\ell_{a;b}m^am^b &\lambda=n_{a;b}m^{*a}m^{*b}\cr
\tau&=-\ell_{a;b}m^an^b & \pi=n_{a;b}m^{*a}\ell^b\\
\epsilon&=-\dfrac{1}{2}(\ell_{a;b}n^a\ell^b-m_{a;b}m^{*a}\ell^b)& \gamma=-\dfrac{1}{2}(\ell_{a;b}n^an^b-m_{a;b}m^{*a}n^b)\cr
\alpha&=-\dfrac{1}{2}(\ell_{a;b}n^am^{*b}-m_{a;b}m^{*a}m^{*b})& \beta=-\dfrac{1}{2}(\ell_{a;b}n^am^b-m_{a;b}m^{*a}m^b)\,.\nonumber
\end{align}
Teukolsky method derives a master equation for the Weyl scalars constructed using the Weyl tensor and the null vectors \cite{No-dynamics-1}. It was shown that these scalars contain useful information about the metric and electromagnetic perturbations.
\paragraph{Hertz Potential.} In our problem we need to know the exact form of metric (and gauge field) perturbations. It was shown in \cite{Kegeles:1979an, Chrzanowski:1975wv, Wald:1978vm} (see \cite{Price:2007} for a review) how to construct field perturbations using a gauge invariant scalar, called the Hertz potential $\Psi_{H}$ which is a solution of Teukolsky master equation. Given the Hertz potential one can construct field perturbations in a specific gauge called \textit{ingoing radiation gauge} (IRG). For this gauge, the Hertz potential for gravitational and Maxwell field perturbations is a solution of Teukolsky master equation with spin $s=-2$ and $s=-1$ respectively. The construction of field perturbations is explicitly

\small{
\begin{align}\label{Hertz map}
h_{\mu\nu}&=\Big( \ell_{(\mu}m_{\nu)}\left[(D+3\epsilon+\bar{\epsilon}-\rho+\bar{\rho})(\delta+4\beta+3\tau)+(\delta+3\beta-\bar{\alpha}-\tau-\bar{\pi})(D+4\epsilon+3\rho)\right] \cr
&-\ell_\mu\ell_\nu(\delta+3\beta+\bar{\alpha}-\bar{\tau})(\delta+4\beta+3\tau)-m_\mu m_\nu(D+3\epsilon-\bar{\epsilon}-\rho)(D+4\epsilon+3\rho)\Big)\Psi_g(x)+c.c,\nonumber\\
\;\delta A_\mu&=\Big(\ell_\mu (\delta+2\beta+\tau)+m_\mu(D+2\epsilon+\rho) \Big)\Psi_A(x)\,.
\end{align}}
\normalsize
\setlength{\baselineskip}{1.1 \baselineskip}
Indeed the Hertz map \eqref{Hertz map} is a map from the solutions of the Teukolsky equation, to the solutions of the linearized field equations for metric (or gauge field) perturbations.  Now the question is whether all solutions of the linearized field equations can be constructed using the Hertz map.

For the case of Kerr black hole, Wald proved \cite{Wald} that there are only specific type of perturbations that lie out of this procedure.  Assuming some regularity conditions, he showed that they are restricted to perturbations to a nearby Kerr black hole with slightly different parameters. In the terminology we used in section \ref{sec uniqueness}, the only solutions to the linearized field equations that cannot be reproduced with the Hertz map are \emph{parametric perturbations} $\de\Phi$. They are perturbations preserving the type D property of the geometry to first order \cite{Wald} (see also section 4 of \cite{No-dynamics-1}). Noticing the argument given in \cite{Durkee:2010qu}, we assume that this is also the case for NHEK geometry, i.e. the only solutions that cannot be constructed using the Hetrz map are parametric perturbations $\de\Phi$.

Therefore the outline of the proof is as follows: As we discussed the solutions to the linearized field equations can be divided into two sets: (I) Those corresponding to a Hertz potential, and (II) parametric perturbations.
The next step is to show that the master equation for the Hertz potential has no solution with the conditions given in the definition \ref{constraints}, i.e. no member in set (I) has our desired conditions. On the other hand, since we showed in the opening of section \ref{sec NHEG parametric} that parametric perturbations satisfy the conditions of definition \ref{constraints}, then we have shown that the only solution with these conditions are parametric perturbations.

Note that the Hertz map formalism is generically developed in the case of vacuum background, therefore we assume in this appendix that the background is a \textit{vacuum} NHEG. This assumption is also necessary in the Kaluza-Klein reduction used in solving the Teukolsky equation in the following. However we did not need this assumption in the arguments of section \ref{sec uniqueness}.

In the following we only need to show that the master equation governing the Hertz potential has no solution compatible with our conditions. It is shown \cite{No-dynamics-1} that the master equation for the Hertz potential corresponding to metric perturbations is the Teukolsky equation with spin $s=-2$, and the master equation for the Hertz potential corresponding to gauge field perturbations is the Teukolsky equation with spin $s=-1$.

It was shown \cite{Durkee:2010ea} that the equation of motion of Teukolsky scalar $\Psi$ over a vacuum NHEG can be Kaluza-Klein reduced to $AdS_2$ space. More precisely, the field equation for the separable ansatz $\Psi=R(t,r)Y(\theta,\varphi^i)$ reduces to the equation of a charged massive scalar $R(t,r)$ over the $AdS_2$ with a homogeneous electric  field, and an eigenvalue equation for $Y$ over the compact surface $H$ (covered by coordinates $\theta,\varphi^i$).  Noting  isometries of the background, one can further expand the solution in eigenstates of $\frac{\partial}{\partial t},\frac{\partial}{\partial \varphi^i} $:
\begin{align}
\Psi=e^{i(\omega t-m_i\varphi^i)}R_{\omega,\lambda,m}(t,r)Y_{\lambda,m_i}(\theta),
\end{align}
and the corresponding field equations become
\begin{align}
\label{R-eq}\left( D^2-\mu^2\right)R(r)&=0\,,\\
\label{Y-eq}\mathcal{O}^{(s)}Y_{\lambda,m_i}(\theta)&=\lambda Y_{\lambda,m_i} (\theta)\,,
\end{align}
where
\be\begin{split}
D_{\mu}&=\nabla_{\mu}-iqA_{\mu},\hspace{1cm}\mu\in \{t,r\}\\
\mu^{2}&=\lambda+q^{2},\\
q&=k^{i}m_{i}-is\,,
\end{split}\ee
and $s$ is the spin of perturbations ($-2$ for gravitational perturbations and $-1$ for Maxwell perturbations). The solutions to \eqref{R-eq} are \textit{hypergeometric} functions \cite{No-dynamics-1,Amsel:2009pu}. The value of $\lambda$ is constrained by the equation on compact space $H$. The regularity of solutions at poles restricts the eigenvalues $\lambda$ to discrete values with a lower bound depending on the field spin. It is shown that the operator $\mathcal{O}^{(s)}$ is self adjoint, therefore its eigenvalues are real.  The most general solution is hence
 \begin{align}
\Psi=\sum_{\omega,\lambda,m_i}e^{i(\omega t-m_i\varphi^i)}R_{\omega,\lambda,m}(t,r)Y_{\lambda,m_i}(\theta)\,.
\end{align}
Since we assume that perturbations are stationary and axisymmetric,  solution reduces to
\begin{align}
\Psi=\sum_{0,\lambda,0}R_{0,\lambda,0}(r)Y_{\lambda,0}(\theta)\,.
\end{align}
According to \eqref{Hertz map}, requiring $h_{\mu\nu}$ to be symmetric under $\xi_2$ exactly fixes the $r$ dependence of $\Psi$ to be
\begin{align}
\Psi_H&=C(\theta)r^2\,.
\end{align}
However, such a radial behavior cannot be constructed using the hypergeometric functions. At large $r$, hypergeometric functions have the asymptotic behavior
\begin{align}\label{R-expansion}
R_\lambda(r)&=\sum_{\lambda}\left( c^{+}_\lambda (r^{\frac{3+ \sqrt{1+4\lambda}}{2}} +\text{subleading})+c^{-}_\lambda (r^{\frac{3- \sqrt{1+4\lambda}}{2}}+\text{subleading}) \right)\,.
\end{align}
Equating $R=r^2$ yields:
\be\begin{split}
c^+_\lambda&=0, \hspace{1cm}\lambda>0\\
c^+_0&=1\,.
\end{split}\ee
The regularity of the eigenstates of \eqref{Y-eq}, however, implies that $\lambda>0$ and therefore a nonvanishing $c^+_0$ leads to a divergent behavior of the angular part.\footnote{
For the NHEK geometry, the solutions to \eqref{Y-eq} are spin-weighted spheroidal harmonics analyzed in \cite{Teukolsky:1973ha}. It turns out that the eigenvalues are
\begin{align}\label{lambda-constraint}
\lambda&=l(l+1),\hspace{1cm}  l\geq Max\{|s|,|m|\}
\end{align}
therefore the condition $\lambda\geq\lambda_{min}>0$ is justified for NHEK space. The positivity of eigenvalues $\lambda$ also hold in NHEK-AdS geometry \cite{Dias:2012pp} and near horizon geometry of cohomogeneity-1 extremal Myers Perry black holes \cite{Durkee:2010ea}. One can argue that the positivity of $\lambda$ is strongly related to the stability of NHEG geometry. In other words, our NHEG perturbation uniqueness holds  for stable NHEG geometries.
} Therefore, a smooth perturbation with the radial behavior $R=r^2$ cannot be constructed, i.e. no perturbation with the specified symmetries can be constructed using the Hertz map. Hence, the only perturbation with our conditions are the perturbation missed by the Hertz map, i.e. the parametric perturbations $\de g_{\mu\nu}$.

\paragraph{Gauge field perturbations.}
The Hertz map for constructing gauge perturbation is
\begin{align}
\delta A_\mu&=\left(\ell_\mu (\delta+2\beta+\tau)+m_\mu(D+2\epsilon+\rho) \right)\Phi_0\,,
\end{align}
imposing the $\xi_2$ symmetry yields $\Phi_0=rF(\theta)$. Therefore according to \eqref{R-expansion} it requires
\be\begin{split}
c^+_\lambda&=0, \hspace{1cm}\lambda\geq 0\\
c^-_{\lambda=0}&=1\,,
\end{split}\ee
which is again violating the regularity constraint $\lambda>0$ which implies $c^-_{\lambda=0}=0$.  Therefore no perturbation with the specified symmetries can be constructed using the Hertz map and  and the only perturbation with our conditions are the parametric perturbations $\de A_{\mu}$ with a variation in NHEG charge $\de q$.
\paragraph{Dilaton field.}
Assuming the symmetry conditions \ref{condition1}, \ref{condition2} implies that a scalar can only depend on the polar coordinate $\theta$. Now for a vacuum background, one can directly solve the linearized field equation, which results that the scalar should be constant allover the spacetime. Remembering the shift symmetry in dilaton field, it is clear that this solution is equivalent to the parametric perturbation of dilaton field.
\paragraph{Generalizations.}
For a generic NHEG, there are some steps to be passed in order to prove the uniqueness theorem. Some of these step are not present in the literature and filling these gaps are not in the scope of this work. However, we give the outline and leave them as conjectures.
\begin{enumerate}
\item Teukolsky equations are written for $d=4$. In higher dimensions, a generalization of Teukolsky formalism is presented in \cite{Durkee:2010ea}. It is shown that requiring the equations to be decoupled, it is not sufficient that the space is Petrov type D (the concept of Petrov classification is extended to $d>4$ in \cite{Coley:2004jv} (and reviewed in \cite{Coley:2007tp}). Moreover the space is required to have a null geodesic congruence with vanishing expansion, shear and torsion. Such a space is called a \textit{Kundt} space. Fortunately NHEG is an example of Kundt spacetimes. In \cite{Godazgar:2011sn}, the Hertz map for gravitational and gauge field perturbation of a higher dimensional Kundt background was given using the decoupled equation of Reall \cite{Durkee:2010qu}.
\item In a more general NHEG, one should prove the positivity of $\mathcal{O}^{(s)}$. No general argument still exists, and this is shown to be valid for different examples individually. Since the positivity of $\mathcal{O}^{(s)}$ is related to the stability of the corresponding NHEG, therefore we expect that this argument is valid only for a stable NHEG, which seems reasonable.
\item It should be checked that the only regular perturbation which is missed in the Hetrz map are parametric perturbations $\de\Phi$. This is only proved for Kerr geometry \cite{Wald}. However an argument is given in \cite{Durkee:2010qu} that this result may extend to NHEG geometries. This is also a gap in the literature.
\item In this appendix, we assumed that the background NHEG is a vacuum solution. We needed this assumption in the construction of field perturbations using the Hertz potential, as well as in the Kaluza-Klein reduction of Teukolsky equation to $AdS_2$ space. However, in the arguments of section \ref{sec uniqueness} we did not need such condition. This suggests that the arguments of this appendix can be generalized to the case of backgrounds containing matter.
\end{enumerate}

\end{document}